\documentclass[aps,prb,final,twocolumn,letterpaper]{revtex4}

\usepackage{graphicx}                            
\usepackage{epstopdf}
\usepackage{amsmath}
\usepackage{physics}
\usepackage{bm} 
\usepackage{amssymb}
\usepackage{hyperref}
\usepackage{xcolor}
\usepackage{array}
\usepackage{multirow}
\usepackage{lipsum}
\usepackage[T1]{fontenc}
\usepackage{multirow}
\usepackage[english]{babel}

\begin{document}

\title{Toward a  global phase diagram of the fractional quantum anomalous Hall effect}


\author{Aidan P. Reddy}
\author{Liang Fu}
\affiliation{Department of Physics, Massachusetts Institute of Technology, Cambridge, Massachusetts 02139, USA}
\begin{abstract}
    Recent experiments on the twisted semiconductor bilayer system $t$MoTe$_2$ have observed integer and fractional quantum anomalous Hall effects, which occur in topological moir\'e bands at zero magnetic field. Here, we present a global phase diagram of $t$MoTe$_2$ throughout the filling range $0< n\leq 1$ substantiated by exact diagonalization calculations. At a magic angle, we find that the system resembles the lowest Landau level (LLL) to a remarkable degree, exhibiting an abundance of incompressible fractional quantum anomalous Hall states and compressible anomalous composite Fermi liquid states. Away from the magic angle, particle-hole symmetry is strongly broken. Some LLL-like features remain robust near half-filling, while others are replaced, predominantly by charge density waves near $n=0$ and anomalous Hall Fermi liquids near $n=1$. Among LLL-like phases, we find the anomalous composite Fermi liquid at $n=\frac{1}{2}$ to be most robust against deviations from the magic angle. Within the band-projected model, we show that strong particle-hole asymmetry above the magic angle 
    results from interaction-enhanced quasiparticle dispersion 
    near $n=1$. Our work sets the stage for future exploration of LLL-like and beyond-LLL phases in fractional quantum anomalous Hall systems. 
    
\end{abstract}

\maketitle

\section{Introduction}

Twisted transition metal dichalcogenide homobilayers ($t$TMDs) host topological moir\'e bands that exhibit spin-valley locking and spin/valley contrasting Chern numbers \cite{wu2019topological}. Owing to band topology and narrow bandwidth,   
small-twist-angle bilayer MoTe$_2$ and WSe$_2$ are predicted to support integer and fractional quantum anomalous Hall (QAH) states \cite{devakul2021magic,li2021spontaneous,crepel2023anomalous,morales2023pressure}. These are chiral topological states that {\it spontaneously} break time-reversal symmetry and exhibit integer and fractionally quantized anomalous Hall conductance $\sigma_{\rm AH}= C e^2/h$ at zero magnetic field respectively.  Recently, electronic compressibility measurements on $t$WSe$_2$ have shown incompressible states with $C=1$ persisting down to zero magnetic field at filling factors $n=1$ and $3$, providing evidence for integer QAH states.
In $t$MoTe$_2$, optically detected Landau fan diagrams reveal signatures of  integer as well as fractional QAH states, with 
$C=-1$,  $-\frac{2}{3}$ and $-\frac{3}{5}$ respectively \cite{cai2023signatures,zeng2023thermodynamic}. 

Very recently, for the first time, the fractionally quantized anomalous Hall effect was observed through transport measurements on $t$MoTe$_2$ \cite{park2023observation}. This observation provides convincing evidence of
a topological phase hosting fractionally charged quasiparticles at zero magnetic field, opening a new frontier in topological physics and quantum materials research. Another group also reported the observation of an integer QAH effect as well as a nearly fractionally quantized anomalous Hall effect in $t$MoTe$_2$ \cite{xu2023observation}.  

Following the experimental breakthrough, recent theoretical works have studied various FQAH states in $t$MoTe$_2$ at specific odd-denominator filling fractions \cite{reddy2023fractional,wang2023fractional,qiu2023interaction,goldman2023zero}. 
Using numerical exact diagonalization, our recent work \cite{reddy2023fractional} showed that  the FQAH state at $n=\frac{2}{3}$ appears robustly over a broad range of twist angles. 
In contrast,  the FQAH state at $n=\frac{1}{3}$ competes with an insulating charge density wave state 
that is favored at experimentally studied twist angles $\theta > \sim 2.3^\circ$.  These conclusions are strongly supported by recent experiments (see below) \cite{zeng2023thermodynamic}.

The formation of FQAH states in $t$TMDs results from (1) exchange-induced {\it spontaneous} spin/valley polarization 
and (2)  strong correlation in spin/valley polarized  Chern bands at partial filling. Broadly speaking, when a Chern band is sufficiently flat \cite{tang2011high, neupert2011fractional,sheng2011fractional,sun2011nearly,regnault2011fractional,song2015topological,zhang2019nearly}, its wavefunctions 
sufficiently resemble those of the lowest Landau level (LLL)\cite{kol1993fractional,qi2011generic,roy2014band,claassen2015position,tarnopolsky2019origin,ledwith2020fractional,wang2021exact,ledwith2022vortexability,paul2023giant,dong2022dirac,wang2023origin,morales2023magic}, and the topological gap to higher bands is sufficiently large, the system can be approximately mapped to
a partially filled Landau level through the band-projected Hamiltonian. 
If this mapping holds faithfully, the existence of FQAH states follows straightforwardly from the well-known fractional quantum Hall states in Landau levels. The key question is: to what extent is the phase diagram of $t$MoTe$_2$ as a function of band filling similar to that of the LLL
\cite{kivelson1992global}? And in what aspect is it fundamentally different?

\begin{figure*}
    \centering
\includegraphics[width=\textwidth]{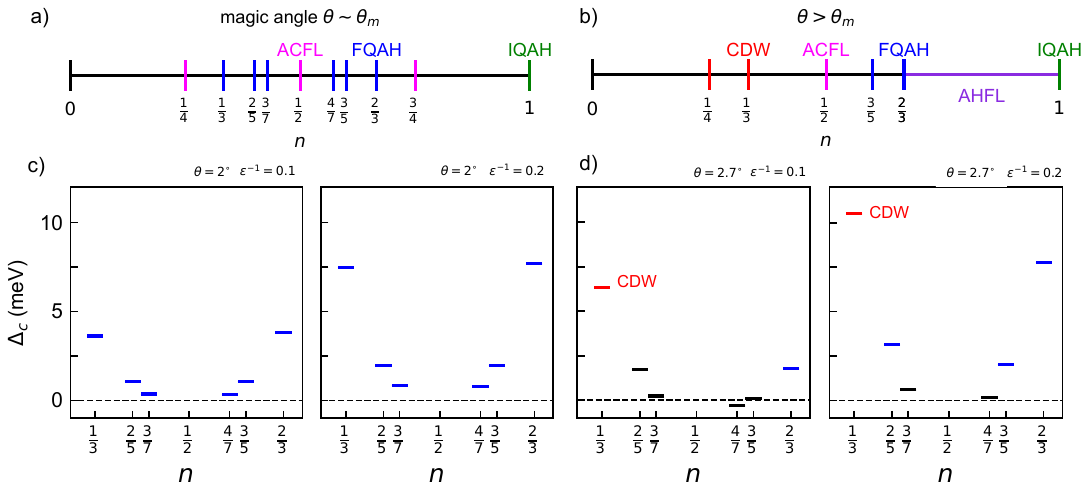}
    \caption{Schematic phase diagrams of $t$MoTe$_2$ with respect to $n$, the number of holes per moiré unit cell, at angles near (a) and greater than (b) the magic angle $\theta_m$. (c,d) Charge gap at several Jain sequence filling fractions at corresponding representative twist angles and two interaction strengths. In (c,d), blue denotes FQAH, red denotes CDW, and black denotes undetermined. Data for fractions with denominators $3$, $5$, and $7$ are obtained from clusters with 27, 25, and 28 unit cells respectively (see Supplemental Material). (F/IQAH: fractional/integer quantum anomalous Hall, ACFL: anomalous composite Fermi liquid, CDW: charge density wave, AHFL: anomalous Hall Fermi liquid.)} 
\label{fig:phasediagram}
\end{figure*}

In this work, we map out the global phase diagram for $t$TMDs throughout the filling range $n\leq 1$. We find that, in the vicinity of a ``magic angle'', an abundance of FQAH states appear and the phase diagram is nearly symmetric about $n=\frac{1}{2}$, which hosts an anomalous composite Fermi liquid state (ACFL) \cite{goldman2023zero,dong2023composite}, see Fig. \ref{fig:phasediagram}(a). ACFL states may also appear at $n=\frac{1}{4}$, $\frac{3}{4}$. In these regards, the system closely resembles the LLL. At larger twist angles, particle-hole symmetry is strongly broken, leading to a coexistence of phases that are familiar with others that are foreign to the LLL. The $n=\frac{1}{2}$ ACFL states and some FQAH states survive beyond the magic angle. In contrast, charge density waves appear at low fillings $n=\frac{1}{3}, \frac{1}{4}$ and a time-reversal-breaking Fermi liquid phase appears in the filling factor range $\frac{2}{3}<n<1$, see Fig. \ref{fig:phasediagram}(b). 
Our findings demonstrate the remarkable robustness of ACFL state at $n=\frac{1}{2}$ with respect to twist angle and establish its central role as the parent 
of adjacent FQAH states in the phase diagram. Our work also reveals an ``anomalous Hall Fermi liquid'' phase  that exhibits an unquantized, filling-dependent anomalous Hall conductivity and has no counterpart in the LLL.    
We discuss the origin of similarities and differences between the  many-body phase diagram of topological bands in $t$TMDs and that of the LLL. The observable consequences of our phase diagram for fractional quantum anomalous Hall systems are also described and compared to recent experiments on $t$MoTe$_2$.

\section{Charge gap phase diagram}  

The continuum model for AA-stacked, K-valley TMD homobilayers describes  holes in a moiré-periodic scalar potential and a layer-pseudospin ``Zeeman'' field that carries a skyrmion texture \cite{wu2019topological}.
In the vicinity of a magic angle, several LLL-like features appear in the lowest moiré band at the single-particle level: the bandwidth nearly vanishes and the Berry curvature becomes nearly uniform \cite{devakul2021magic}; the deviation from the so-called trace condition is minimized \cite{morales2023pressure,reddy2023fractional,dong2023composite}; and the general bound on the topological band gap is closest to saturation 
\cite{onishi2023quantum}. Based on 
our large-scale density functional theory (DFT) calculations, it is found that $\theta_m\approx 2^{\circ}$ for twisted bilayer MoTe$_2$ \cite{reddy2023fractional} and $\theta_m\approx 1.5^\circ$ for twisted bilayer WSe$_2$ \cite{devakul2021magic}. We note that the use of different DFT parameters results in some variation in the magic angle \cite{wu2019topological,tang2021geometric,wang2023fractional,zhang2023polarization,mao2023lattice,yu2023fractional}.

At the many-body level, our previous exact diagonalization (ED) calculations indeed find that the neutral energy gaps of the $n=\frac{1}{3}$ and $\frac{2}{3}$ FQAH states are both maximized near a magic angle $\theta_m$ \cite{reddy2023fractional}. Here, we present a comprehensive ED study of 
$t$MoTe$_2$ both near and above the magic angle and throughout the filling range $n \leq 1$,  finding a plethora of incompressible and compressible states. Our results are obtained on a finite-sized torus using the continuum model for $t$MoTe$_2$ with the parameters of Ref. \cite{reddy2023fractional} projected to the lowest moiré band and a Coulomb interaction $V(r)=\frac{e^2}{\epsilon r}$. Given that 
previous studies have shown 
the ground state over a wide range of fillings $n\leq 1$ to be fully spin/valley polarized \cite{crepel2023anomalous, reddy2023fractional}, all ED calculations in this work are performed in the 
fully spin/valley polarized sector. Our model and methods are described in the Supplemental Material of Ref. \cite{reddy2023fractional}.

The defining property of an incompressible state is a finite charge gap in the thermodynamic limit. Here we study the charge gap
$\Delta_c(N)=\mu^{+}_{N}-\mu^{-}_{N}$ with $\mu^{\pm}_{N}=\pm(E_{GS}(N\pm1)-E_{GS}(N))$ where $N$ is the number of holes and $E_{GS}(N)$ is the ground state energy at a given $N$ and a fixed finite system geometry. 
The thermodynamic limit of $\Delta_c$ can be measured via compressibility and, in a state with quasiparticles of fractional charge $e^*=\frac{p}{q}$, is related to the transport activation gap $\Delta_t$ as $p\Delta_c = q\Delta_t$ \cite{girvin2002quantum}.
Note that the charge gap is distinct from the neutral energy gap, the difference between the ground and first excited state energies at fixed particle number.

In Fig. \ref{fig:phasediagram}, we show the charge gaps at the Jain sequence filling fractions $n=\frac{p}{2p+1}$ where $p=1,2,3$ and their particle-hole conjugates at two twist angles 
and two interaction strengths. 
At $\theta=2^{\circ}$, representative of the system near the magic angle $\theta_m$, the charge gap is positive at all $p$, decreases with increasing $p$, and is nearly particle-hole symmetric under $n\rightarrow 1-n$. By inspecting the many-body spectra, we confirm that these incompressible states are all FQAH states (see Supplemental Material). Remarkably, the decreasing charge gaps of FQAH states as the filling approaches $\frac{1}{2}$ closely resembles the Jain sequence of fractional quantum Hall states in the LLL, despite being at zero magnetic field.      

In contrast, at $\theta=2.7^{\circ}$, representative of a broad range of angles $\theta >\theta_m$, the charge gap exhibits strong particle-hole asymmetry, see Fig. \ref{fig:phasediagram}(d). 
$n=\frac{1}{3}$ exhibits a large charge gap associated with an insulating charge density wave rather than an FQAH state \cite{reddy2023fractional,morales2023pressure}. On the other hand, the $n=\frac{2}{3}$ state 
is an FQAH state with the topological order of the particle-hole conjugate of the $\frac{1}{3}$ Laughlin state in the LLL. Compared to that at $n=\frac{2}{3}$, other FQAH states are more fragile when $\theta$ exceeds $\theta_m$.  For $\epsilon^{-1}=0.1$, the charge gap at $n=\frac{4}{7}$ become negative and that at $n=\frac{3}{5}$ is positive but very small. When interaction strength increases to $\epsilon^{-1}=0.2$, the charge gaps at both filling fractions increase and are positive. Compared to the FQAH state at $n=\frac{2}{3}$, stronger interactions are necessary to stabilize FQAH states at these filling fractions away from the magic angle. 

In the Supplemental Material, we compare charge gaps at various filling fractions between magic-angle $t$MoTe$_2$ and the LLL on identical finite system geometries, which show a striking similarity at \emph{all} carrier densities. This indicates that the resemblance between the two systems is not limited to specific filling fractions. We also study the evolution with respect to the twist angle $\theta$ of the many-body spectra at several rational filling fractions. 
In addition to the fractions shown here, we find that many others, including $n=\frac{1}{5}$, $\frac{4}{5}$, $\frac{4}{9}$, $\frac{2}{7}$, $\frac{5}{7}$, and $\frac{5}{9}$ may host FQAH states at the magic angle. 

Our work reveals the physical consequences of the magic angle in $t$MoTe$_2$. While the $n=\frac{2}{3}$ FQAH state exists robustly over a wide range of twist angles, only at $\theta \sim 2^{\circ}$ does the many-body phase diagram resemble the LLL at all filling factors. 





\begin{figure}
    \centering
\includegraphics[width=\columnwidth]{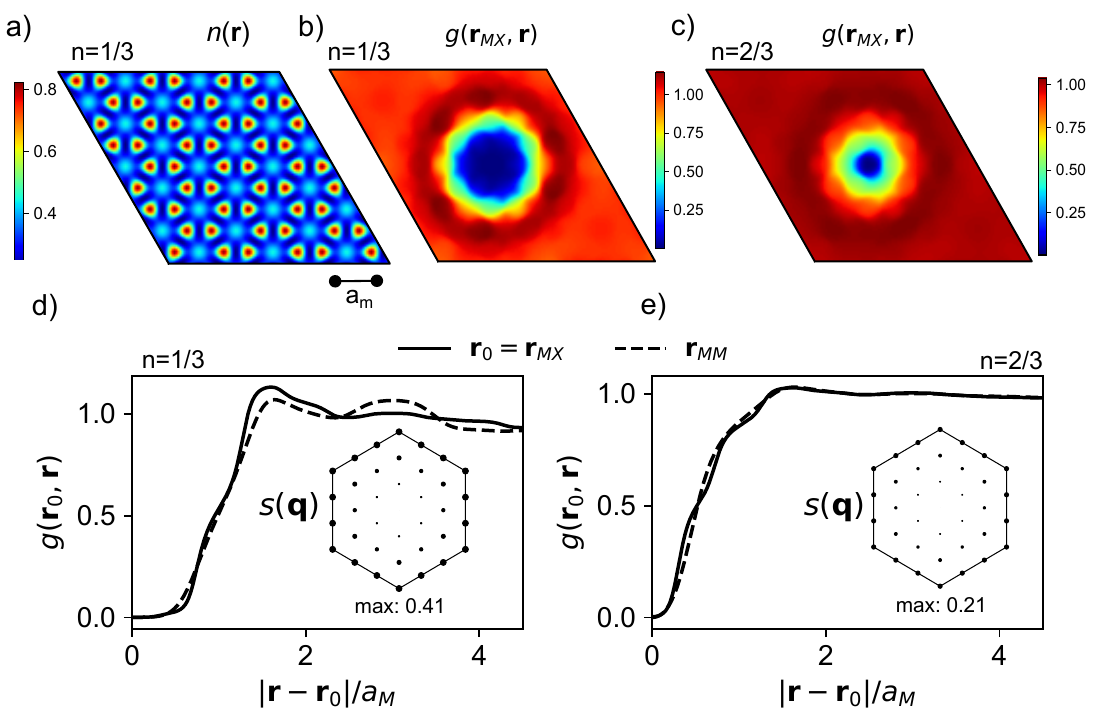}
    \caption{(a) Charge density at $n=\frac{1}{3}$. MM stacking regions appear light blue and MX/XM stacking regions are red. Pair correlation functions $g(\bm{r}_0,\bm{r})$ and projected static structure factors $s(\bm{q})$ of the $n=\frac{1}{3}$ (b,d) and $\frac{2}{3}$ (c,e) FQAH fluids at $\theta=2^{\circ} \sim \theta_m$. Line cuts are taken along $\bm{r}-\bm{r}_0$ going from the center to the bottom right corner of (b,c). Line cuts with $\bm{r}_0$ at two high symmetry points in the moiré unit cell, MX and MM, are both shown. $\epsilon^{-1}=0.2$ and the same system with 27 moiré unit cells is used in all panels.}
\label{fig:FQAHcorr}
\end{figure}

\section{Fractional quantum anomalous Hall fluids}
To 
further characterize the FQAH states, we now study their two-body correlations. In Fig. \ref{fig:FQAHcorr}, we show the pair correlation functions
\begin{align}
    g(\bm{r},\bm{r}')=\frac{\langle n(\bm{r})n(\bm{r}')\rangle -\delta(\bm{r}-\bm{r}')\langle n(\bm{r})\rangle}{\langle n(\bm{r})\rangle \langle n(\bm{r}') \rangle}
\end{align}
and projected density structure factors
\begin{align}
\begin{split}
    s(\bm{q}) = \frac{\langle \bar{\rho}(-\bm{q})\bar{\rho}(\bm{q})\rangle}{N} - \delta_{\bm{q},\bm{0}}
\end{split}
\end{align}
of the $n=\frac{1}{3}$, $\frac{2}{3}$ FQAH states near the magic angle. Here $\bar{\rho}(\bm{q}) = \mathcal{P}\sum_{i}e^{-i\bm{q}\cdot\bm{r}_i}\mathcal{P}$ is the projected density fluctuation operator where $\mathcal{P}$ is the projector onto the lowest moiré band and $i$ indexes holes. $\langle\rangle$ denotes an expectation value with respect to the many-body ground state(s). 
The local density operator is $n(\bm{r}) = \sum_{l}\psi^{\dag}_{l}(\bm{r})\psi_{l}(\bm{r})$ where $\psi^{\dag}_{l}(\bm{r})$ creates a hole at position $\bm{r}$ in layer $l$.

The pair correlation functions of the $n=\frac{1}{3}$ and $\frac{2}{3}$ states show clear exchange-correlation holes at small separation and remain near $1$ at large separation. At $n=\frac{1}{3}$, the pair correlation function oscillates at intermediate range, indicating incipient crystalline order. The projected structure factors 
do exhibit some non-uniformity but lack the strong peaks expected of a crystal. 
Both of these correlation functions exhibit characteristics of a fluid with strong short-range correlations. We remark that the presence of the layer degree of freedom means that, in general, $g(\bm{r},\bm{r})\neq 0$ even when spin/valley is fully polarized. Additional analysis of FQAH correlation functions, including at $n=\frac{2}{5}$, $\frac{3}{5}$ is presented in the Supplemental Material.


The charge gap phase diagram (Fig. \ref{fig:phasediagram}) shows that, throughout a wide range of twist angles, the system, hosts a sequence of FQAH states analogous to the Jain sequence in the LLL. Having studied these incompressible topological fluids, we now turn to the filling fraction $n=\frac{1}{2}$ about which they are centered.


\section{Anomalous composite Fermi liquid at Half Filling} 

\begin{figure}
    \centering
\includegraphics[width=\columnwidth]{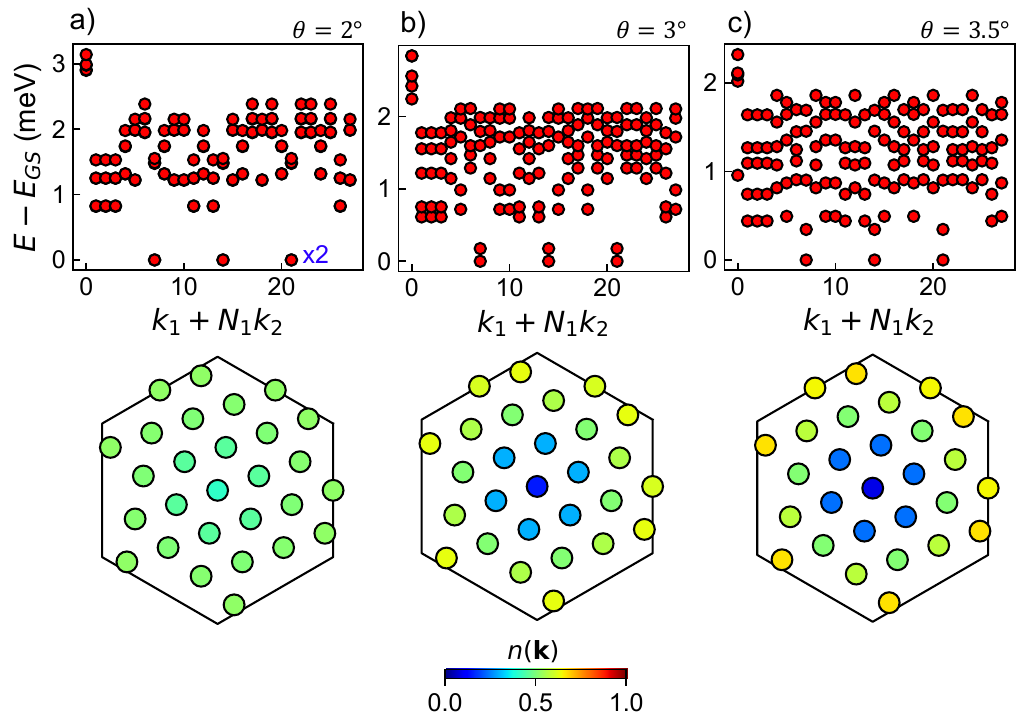}
    \caption{Low-lying many-body energy spectra (top) and ground state momentum distribution function $n(\bm{k})$ (bottom) at $n=\frac{1}{2}$ and several twist angles. In each case $\epsilon^{-1}=0.1$ and the lowest 5 energy levels in each momentum sector are shown. The system size is 28 moiré unit cells.}
\label{fig:ACFLED}
\end{figure}

Previously, we showed that metallic states analogous to the composite Fermi liquids of the LLL but at zero magnetic field exist at $n=\frac{1}{2}$ and $n=\frac{3}{4}$,  which we dubbed ``anomalous composite Fermi liquids'' (ACFL) \cite{goldman2023zero}. Here, we extend our earlier study to a larger system (with 28 moiré unit cells) and a broader range of angles beyond $\theta=2^{\circ}$. Fig. \ref{fig:ACFLED} shows the many-body energy spectra at $n=\frac{1}{2}$ for three twist angles $\theta=2^{\circ}$, $3^{\circ}$, and $3.5^{\circ}$. Near the magic angle, the ground states come in quasi-degenerate pairs as they do in the half-filled LLL (where they are related by center-of-mass magnetic translations \cite{haldane1985finite,girvin2019modern}), and their many-body momenta also match those of the half-filled LLL on a torus of identical geometry, showing that the $n=\frac{1}{2}$ state is a composite Fermi liquid. Unlike incompressible candidate competing states, the composite Fermi liquid's finite-size ground state degeneracy is sensitive to the system geometry and can be interpreted from trial wave functions as arising from compact configurations of the composite Fermi sea \cite{rezayi1994fermi,jain1997quantitative,rezayi2000incompressible,geraedts2018berry,wang2019lattice,wang2021exact}.

As the twist angle increases from $2^{\circ}$ to $3.5^{\circ}$,  no ground state level crossing occurs, suggesting that the system at $n=\frac{1}{2}$ remains in the same composite Fermi liquid phase throughout. 
In contrast, with the same interaction strength $\epsilon^{-1}=0.1$, the FQAH state at $n=\frac{2}{3}$ undergoes a phase transition at $\theta\approx 3^{\circ}$  where a level crossing between the FQAH ground state manifold and excited states occurs \cite{reddy2023fractional}. 
This observation suggests that the ACFL state at $n=\frac{1}{2}$ is more resilient against departure from the magic angle than the FQAH state at $n=\frac{2}{3}$. In the Supplemental Material, we provide similar evidence for the $n=\frac{1}{4}, \frac{3}{4}$ ACFL states near the magic angle, and show that, as twist angle increases, both these ACFL states undergo level-crossing transitions before the ACFL state at $n=\frac{1}{2}$.

We also calculate the momentum distribution function $n(\bm{k})=\frac{1}{N_{GS}}\sum_{i}\bra{\Psi_i}c^{\dag}_{\bm{k}}c_{\bm{k}}\ket{\Psi_i}$ (where $c^{\dag}_{\bm{k}}$ creates a hole in a moiré Bloch state and $i$ runs over a set of $N_{GS}$ degenerate many-body ground states), shown in Fig. \ref{fig:ACFLED}. At $\theta=2^{\circ}$, $n(\bm{k})\approx 0.5$ is nearly constant and, in particular, smooth. This shows that the state is not a Landau Fermi liquid in which $n(\bm{k})$ is necessarily discontinuous at the Fermi surface in the thermodynamic limit. 
At larger angle, $n(\bm{k})$ 
varies throughout the Brillouin zone but still does not exhibit a sharp jump. 
We remark that the momentum distribution function is experimentally accessible through angle-resolved photoemission spectroscopy.

Since $n(\bm{k})\rightarrow 1-n(-\bm{k})$ under a particle-hole transformation, the  ground state wavefunction of ACFL state beyond the magic angle is clearly {\it not} invariant under the particle-hole transformation. (Whether there exists an ``emergent'' particle-hole symmetry at low energy is an open question.) At $\theta=3.5^{\circ}$, the spectrum is on the verge of a level crossing between the higher-energy partner of the quasi-degenerate ACFL ground states, indicating a phase transition out of the ACFL. The nature of the many-body ground state following this transition is an interesting open question for future work. 


So far we have seen that (1) near a magic angle, the system resembles the LLL to a remarkable extent; and (2) above this magic angle, some LLL-like features remain in the neighborhood of half filling, while other new phases appear at fillings away from half filling. The similarity of our many-body phase diagram at the magic angle with the LLL supports the recent argument that the wavefunctions of the magic-angle moir\'e band can be approximately mapped to those of the LLL \cite{morales2023magic}. Equally important is the departure from the LLL analog at larger twist angles, to which we now turn our attention.



\section{Microscopic origin of particle-hole asymmetry} 

A generic band-projected Hamiltonian takes the form
\begin{align}\label{eq:particleham}
\begin{split}
    H &= \sum_{\bm k}\varepsilon(\bm{k}) c^{\dag}_{\bm k}c_{\bm k} + \frac{1}{2} \sum_{\bm{k}\bm{p}\bm{q}} V_{(\bm{k}+\bm{q})(\bm{p}-\bm{q})\bm{k}\bm{p}}c^{\dag}_{\bm{k}+\bm{q}}c^{\dag}_{\bm{p}-\bm{q}}c_{\bm{p}}c_{\bm{k}}
\end{split}
\end{align}
where 
spin is neglected. Therefore, it is determined entirely by the dispersion $\varepsilon(\bm{k})$ and the interaction matrix elements $V_{\bm{k}'\bm{p}'\bm{k}\bm{p}}$. The latter are determined in turn by the band's single-particle wavefunctions and the two-body interaction potential. Any deviation between the phase diagrams of given band-projected model and the LLL originates from a deviation in these features.

Motivated by our observation of strong asymmetry between $n$ and $1-n$ states above the magic angle as shown in Fig. \ref{fig:phasediagram}, we now consider the 
band-projected Hamiltonian under the particle-hole transformation.
Under the particle-hole transformation $d^{\dag}(\bm{r})=c(\bm{r})$ 
or, equivalently, $d^{\dag}_{\bm k}=c_{-\bm k}$, where $c_{\bm k}$ 
annihilates the particle in Bloch state $\ket{\bm{k}}$, Eq. \ref{eq:particleham}
can be rewritten in terms of hole operators as
\begin{align}\label{eq:holeham}
\begin{split}
 H 
 &= \sum_{{\bm k}} \tilde{\varepsilon}(\bm k) d^{\dag}_{{\bm k}}d_{{\bm k}} +  \frac{1}{2}\sum_{\bm{k}\bm{p}\bm{q}} \tilde{V}_{(\bm{k}+\bm{q})(\bm{p}-\bm{q})\bm{k}\bm{p}}d^{\dag}_{\bm{k}+\bm{q}}d^{\dag}_{\bm{p}-\bm{q}}d_{\bm{p}}d_{\bm{k}}
\end{split}
\end{align}
where we neglect a constant energy shift. The full-band Slater determinant state $\ket{\Psi_f}=(\prod_{\bm{k}}c^{\dag}_{\bm{k}})\ket{0}$ is the vacuum for holes, $d_{\bm{k}}\ket{\Psi_f}=0$. The interaction matrix elements of holes 
are related to 
those of electrons as
\begin{align}
    \begin{split}
        \tilde{V}_{\bm{k}'\bm{p}'\bm{k}\bm{p}} = V^*_{-\bm{k}'-\bm{p}'-\bm{k}-\bm{p}}.
    \end{split}
\end{align}
$\tilde{\varepsilon}(\bm{k})$ describes the energy-momentum dispersion of  a single particle removed from an otherwise full band,
\begin{align}
    \begin{split}\label{eq:holedispersion}
        \tilde{\varepsilon}(\bm{k}) &= -\varepsilon(-\bm{k}) - \Sigma(-\bm{k}). 
    \end{split}
\end{align}
Here, $\Sigma(\bm{k})$ is a 
self-energy term coming from the interaction between the 
electron at $\bm{k}$ and all others in the full 
band (see Supplemental Material for definition).
In the lowest Landau level, 
both $\varepsilon(\bm{k})$ and $\Sigma(\bm{k})$ are $\bm{k}$-independent, producing particle-hole symmetry.
More generally, applying a time reversal transformation $H\rightarrow \mathcal{T}H\mathcal{T}^{-1}$ (where $\mathcal{T}d_{\bm{k}}\mathcal{T}^{-1}=d_{-\bm{k}}$ and $\mathcal{T} i \mathcal{T}^{-1}=-i$) to Eq. \ref{eq:holeham} and comparing to Eq. \ref{eq:particleham} shows that particle-hole symmetry arises when the condition $\tilde{\varepsilon}(\bm{k})=\varepsilon(-\bm{k})$ is satisfied. Any particle-hole asymmetry between states at filling fractions $n$ and $1-n$ in the phase diagram of a band-projected model has its origin in the inequivalence between the energy dispersion of a particle added to an empty band and that of a particle removed from a filled band.


\begin{figure}
    \centering
\includegraphics[width=\columnwidth]{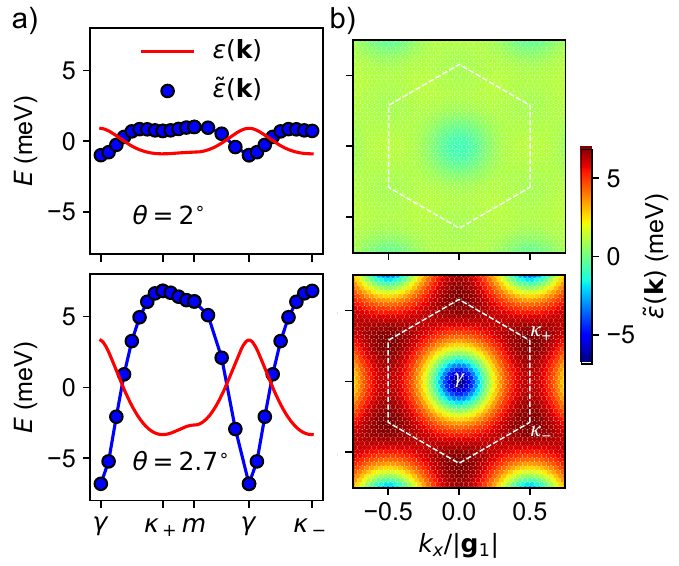}
    \caption{ (a) Comparison of hole dispersion $\varepsilon(\bm{k})$ at $n=0$ and electron quasiparticle dispersion $\tilde{\varepsilon}(\bm{k})$ at $n=1$ of the lowest moiré band assuming full valley polarization at $\theta=2^{\circ}$ and $\theta=2.7^{\circ}$. Note that $\tilde{\varepsilon}(\bm{k})$ depends on the two-body interaction potential for which we use a Coulomb interaction with $\varepsilon^{-1}=0.1$. (b) Color plots of the electron quasiparticle dispersion. All dispersions are shifted by a constant to be centered about zero energy.} 
\label{fig:holeDispersion}
\end{figure}

To shed light on the approximate particle-hole symmetry near the magic angle and lack thereof at larger angles, we now calculate $\tilde\varepsilon(\bm{k})$ in $t$MoTe$_2$ and compare it with the bare energy dispersion $\varepsilon(\bm{k})$. A note on terminology before proceeding.  
Since the moiré band of interest is a valence band, we define the filling factor $n$ as the number of holes (i.e. carrying positive charge) per moiré unit cell relative to charge neutrality. 
These holes are the elementary charge carriers in the moir\'e superlattice. The $n=1$ IQAH state is formed when holes completely fill the topmost moir\'e valence band of one spin/valley. Removing a hole from the $n=1$ state creates a quasiparticle carrying negative charge, which we refer to as an electron quasiparticle. The energy dispersion of a hole at charge neutrality is simply $\varepsilon(\bm{k})$ as determined by the continuum model, whereas the energy dispersion of an electron quasiparticle in the QAH state, $\tilde{\varepsilon}(\bm{k})$, is affected by interaction-induced self-energy as stated above.  

In Fig. \ref{fig:holeDispersion}, we compare the hole $\varepsilon(\bm{k})$ and electron quasiparticle  $\tilde{\varepsilon}(\bm{k})$ dispersions in the lowest $t$MoTe$_2$ moiré band 
at two representative twist angles, $\theta=2.0^{\circ}$ and $2.7^{\circ}$.
At $\theta=2^{\circ}$, the self-energy has an insignificant influence on the electron quasiparticle dispersion so $\tilde{\varepsilon}(\bm{k})\approx -\varepsilon(-\bm{k})$.
On the other hand, at $\theta=2.7^{\circ}$, the self-energy approximately doubles the electron quasiparticle bandwidth. The overall energy scale of interactions is set by $\frac{e^2}{\epsilon a_M}\propto \theta$. At $\theta=2^{\circ}$, $\Delta\Sigma(\bm{k}) \equiv \max(\Sigma(\bm{k}))-\min(\Sigma(\bm{k}))=0.58$ meV and at $\theta=2.7^{\circ}$, $\Delta\Sigma(\bm{k})= 7.0$ meV (given $\epsilon^{-1}=0.1$). In going from $\theta=2^{\circ}$ to $2.7^{\circ}$, the overall interaction energy scale grows by a small factor of $\frac{2.7}{2}=1.35$ while $\Delta\Sigma(\bm{k})$ grows by a much larger factor, $\frac{7.0}{0.58} \approx 12$. Therefore, the $\bm{k}$-dependence of $\Sigma(\bm{k})$ is much stronger above than it is near the magic angle.



The physical origin of the enhanced dispersion for electron quasiparticles above the magic angle is as follows. The self-energy can be decomposed into Hartree and Fock contributions, $\Sigma(\bm{k}) =\Sigma^{H}(\bm{k})+\Sigma^{F}(\bm{k})$. We have directly confirmed that the $\bm{k}$-dependent part of the self-energy above the magic angle is dominated by the Fock term,
which is an interaction-potential-weighted average of quantum distances 
between intra-unit-cell wavefunctions $\ket{u_{\bm{k}}}$ and $\ket{u_{\bm{k}+\bm{q}}}$:
\begin{align}\label{eq:fock}
    \begin{split}
        \Sigma^{F}({\bm k}) 
        &= -\int\frac{d^2q}{(2\pi)^2}V(\bm{q})|\bra{u_{\bm{k}+\bm{q}}}\ket{u_{\bm{k}}}|^2.
    \end{split}
\end{align}
As we showed in Ref. \cite{reddy2023fractional}, the Bloch states in the lowest moiré band are strongly layer polarized except in the vicinity of $\gamma$, where they are strongly layer hybridized. The change in the wavefunctions' layer character at $\gamma$ causes a peak in the Fock self-energy 
that adds constructively with the bare energy dispersion $\varepsilon(\bm{k})$
and thereby enhances the quasiparticle bandwidth at $n=1$.


Our finding that, near the magic angle, the quasiparticle dispersions near $n=0$ $\varepsilon(\bm{k})$ and $n=1$ $\tilde{\varepsilon}(\bm{k})$ are both narrow--and consequently that the condition for particle-hole symmetry $\tilde{\varepsilon}(\bm{k})=\varepsilon(-\bm{k})$ is only weakly violated--is consistent with our finding of approximate particle-hole symmetry in the magic-angle many-body phase diagram (see Fig. \ref{fig:phasediagram}). Above the magic angle, we find in contrast that 1) $\varepsilon(\bm{k})$ is broad to begin with, 2) $\tilde{\varepsilon}(\bm{k})$ is approximately twice as broad due to strong $\bm{k}$-dependence of the Fock energy, and 3) $\varepsilon(\bm{k})$ has two degenerate minima at the corners of the moiré Brillouin zone whereas $\tilde{\varepsilon(\bm{k})}$ has a unique minimum at its center (see Fig. \ref{fig:holeDispersion}).
These properties are consistent with the enhanced-particle hole asymmetry in the above-magic-angle many-body phase diagram shown in Fig. \ref{fig:phasediagram}.

\begin{figure}
    \centering
\includegraphics[width=0.8\columnwidth]{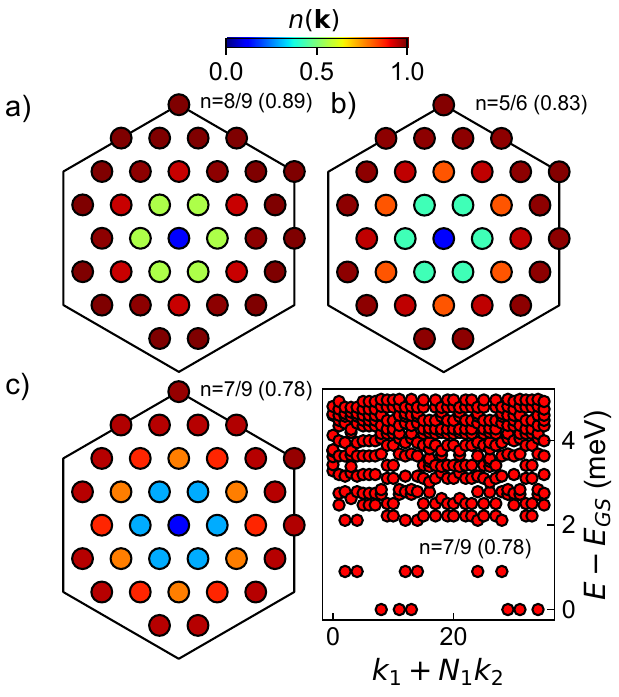}
    \caption{Momentum distribution function (see main text) of the many-body ground states at several carrier densities $\frac{2}{3}< n< 1$ at. $\theta=2.70^{\circ}$, $\epsilon^{-1}=0.1$, and the system includes $36$ moiré unit cells.} 
\label{fig:AHFLED}
\end{figure}

\begin{figure}
    \centering
\includegraphics[width=\columnwidth]{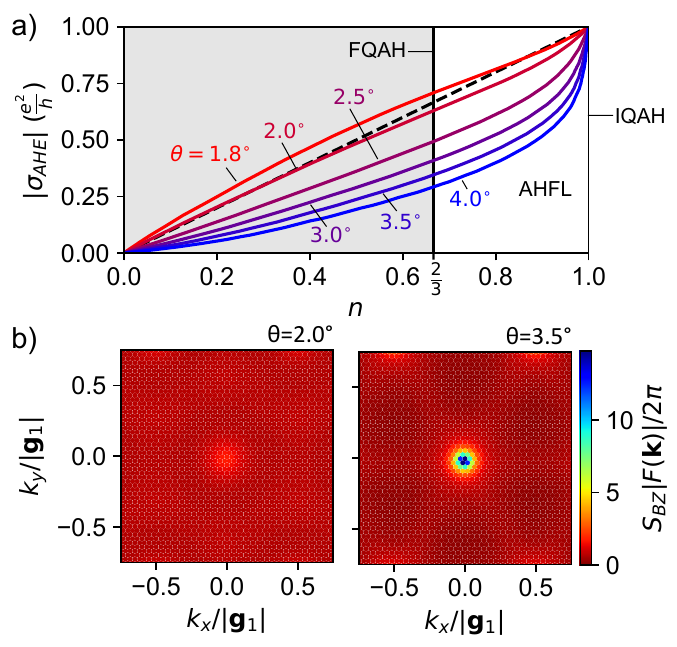}
    \caption{(a) Intrinsic anomalous Hall conductance of a fully  spin/valley polarized Fermi gas in the lowest moiré band as a function of hole density at several twist angles. The white background denotes the region hosting the anomalous Hall Fermi liquid phase studied in this work at angles larger than the magic angle $\theta_m \sim 2^{\circ}$ to which the data shown are applicable. (b) Berry curvature distribution of the lowest moiré band.}
\label{fig:hallconductance}
\end{figure}

\section{Fermi liquid and unquantized anomalous Hall effect} 
With this understanding of particle-hole asymmetry in the band-projected model for $t$MoTe$_2$ above the magic angle, and in particular of the broadened quasiparticle dispersion $\tilde{\varepsilon}(\bm{k})$ in the $n=1$ QAH state, we now study its consequences at finite doping $n=1-\delta$ with $\delta>0$.
For small doping $\delta$, the low-energy physics of our system maps to a uniform electron gas. Provided that the density of electron quasiparticles $\delta$ is not too low, it is natural to expect that its ground state is a Fermi liquid.
The reduction of electron quasiparticle mass by interaction-induced self energy near $n=1$ is also beneficial to the formation of a Fermi liquid \cite{lauchli2013hierarchy}. 
From band-projected ED calculations, we indeed find
a fully spin/valley-polarized, metallic state with a filling-dependent anomalous Hall conductivity in the carrier density range $\frac{2}{3}<n<1$ that we refer to as an \emph{anomalous Hall Fermi liquid}.

Fig. \ref{fig:AHFLED} shows 
the momentum distribution function $n(\bm{k})$ at three filling fractions $n=0.89, 0.83, 0.78$. Unlike the ACFL state at $n=\frac{1}{2}$, at these fillings 
$n(\bm{k})$ drops sharply across a circle centered at $\gamma$,
indicating the presence of a Fermi surface expected from the quasiparticle dispersion $\tilde{\varepsilon}(\bm{k})$. Moreover, the degenerate ground states' many-body momenta match those expected from simply filling electrons according to the quasiparticle dispersion $\tilde{\varepsilon}(\bm{k})$. 
In Fig. \ref{fig:AHFLED}
(d), for instance, the sixfold ground state degeneracy at $n=\frac{7}{9}$ ($\delta=\frac{2}{9}$) comes from adding $36\times \frac{2}{9}=8$ electrons to the 7 Bloch states closest to $\gamma$ and one of the $6$ momenta in the next available shell.  
Similar data obtained from other finite system geometries are also shown in the Supplemental Material. 

In contrast to the Fermi liquid phase close to $n=1$, correlation effects are much stronger at low fillings close to $n=0$ because the effective mass of holes is much larger than that of electron quasiparticles, as seen in Fig. \ref{fig:holeDispersion}. As a consequence, we find that above the magic angle, the $n=\frac{1}{4}$ state is a $2\times 2$ charge density wave (see below), whereas the $n=\frac{3}{4}$ state is a Fermi liquid. Our finding of Fermi liquids stabilized by interaction-enhanced dispersion near unity filling echoes previous studies of other Chern band systems \cite{lauchli2013hierarchy,abouelkomsan2020particle,abouelkomsan2023quantum}.

Having established its existence, we now study the properties of the ferromagnetic Fermi liquid state in $t$MoTe$_2$. In Fig. \ref{fig:hallconductance}, we show the zero-temperature, intrinsic anomalous Hall conductance 
\begin{eqnarray}
\sigma_{AH}=\frac{1}{2\pi}\frac{e^2}{h}\int d^2\bm{k}\,\theta(\varepsilon_F-\varepsilon(\bm{k}))F(\bm{k})
\end{eqnarray} 
where $F(\bm{k})$ is the Berry curvature 
of the lowest moiré band in the presence of full spin/valley polarization. 
We note that this 
this formula is applicable only when the system is in 
a Fermi liquid phase. Near the magic angle, a relatively uniform Berry curvature distribution leads to $\sigma_{AH}\approx n\frac{e^2}{h}$. In contrast, at larger twist angles, the Berry curvature  has a hotspot near the band maximum as shown in Fig. \ref{fig:hallconductance}(b). This leads to a rapid reduction in 
anomalous Hall conductivity $\sigma_{AH}<n\frac{e^2}{h}$ as the filling is reduced from $n=1$ and the system enters the anomalous Hall Fermi liquid phase, before rising to the quantized value $\sigma_{AH} =\frac{2 e^2}{3h}$ in the $n=\frac{2}{3}$ FQAH state. 

In addition to the QAH states at $n=1,\frac{2}{3}, \frac{3}{5}$, the recent transport experiment on $t$MoTe$_2$ reveals intriguing behavior as a function of carrier density \cite{park2023observation}. 
Notably, it is found that (1) $R_{xy}\gg R_{xx}$ throughout the filling range $\frac{1}{2} \leq n \leq \frac{2}{3}$;   
(2) as the band filling increases from $\frac{2}{3}$ to $1$, $R_{xy}$ drops rapidly, is comparable to $R_{xx}$ over an extended filling range, and eventually rises to the quantized value $\frac{h}{e^2}$ at $n=1$. The distinct transport behaviors at $\frac{1}{2} \leq n \leq \frac{2}{3}$ and $\frac{2}{3} \leq n < 1$ agree,  respectively, with the LLL-like FQAH regime and the anomalous Hall Fermi liquid regime of $t$MoTe$_2$ above the magic angle. Our theory then further predicts the existence of CDW states at $n<\frac{1}{2}$, to which we turn our attention now.

\begin{figure}
    \centering
\includegraphics[width=\columnwidth]{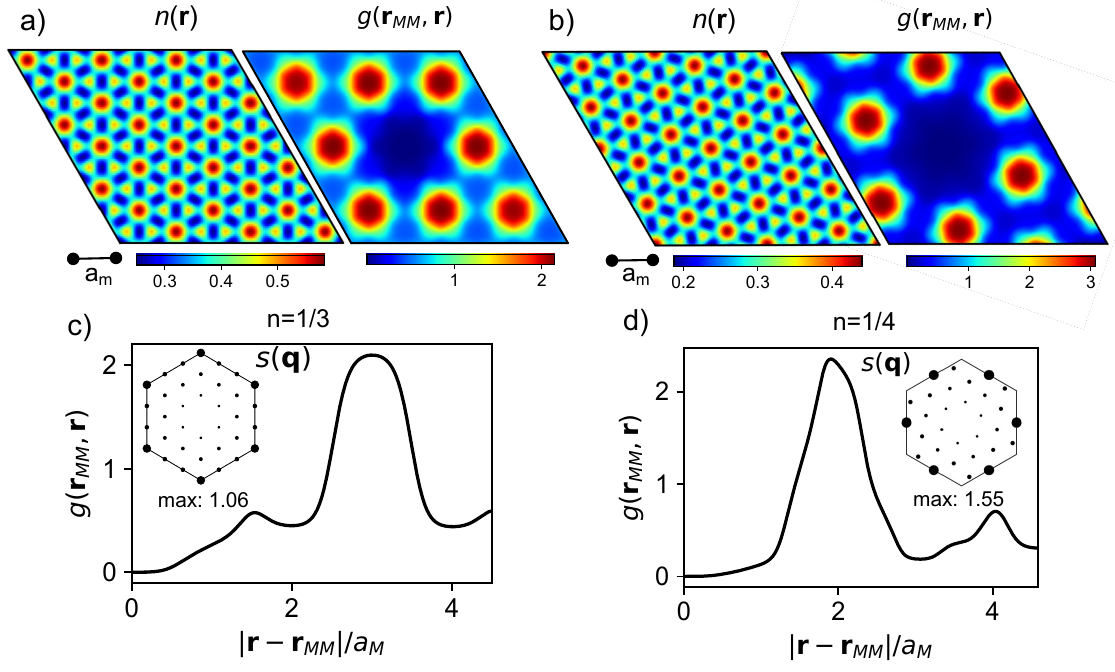}
    \caption{Hole density and pair correlation functions of the charge density wave states appearing above the magic angle at filling factors $n=\frac{1}{3}$ (a) and $\frac{1}{4}$ (b), exhibiting clear crystal structure. (c,d) Line cuts of the pair correlation functions along $\bm{r}-\bm{r}_0$ going from the center to the bottom right corner of (a,b). Insets to (c,d) show the projected structure factors. MM stacking regions are red and MX/XM stacking regions are yellow in the charge density plots of (a,b). Data are from clusters with 27 (a,c) and 28 (b,d) unit cells. $\theta=2.7^{\circ}$ and $\epsilon^{-1}=0.2$.} 
\label{fig:CDW}
\end{figure}

\section{Charge density waves at $n<\frac{1}{2}$}
By explicit calculation of the interaction-renormalized dispersion near $n=1$, we have shown that the band-projected continuum model is strongly particle-hole asymmetric above the magic angle. The model's intrinsic particle-hole asymmetry produces a particle-hole asymmetric diagram with charge density waves at low filling factors in stark contrast to the AHFL phase near $n=1$.
In Fig. \ref{fig:CDW}, we show the pair correlation functions and projected structure factors at $n=\frac{1}{3}$, $\frac{1}{4}$ at $\theta=2.7^{\circ} > \theta_m$. Unlike at smaller angles, the charge density here is peaked at MM stacking sites that form a triangular lattice \cite{reddy2023fractional}. The pair correlation functions exhibit clear crystalline order with $N-1$ distinct peaks 
where $N$ is the total number of holes. Similarly, the projected structure factors exhibit peaks at the moiré Brillouin zone corners and $m$ points, indicating crystalline order with $\sqrt{3}\times \sqrt{3}$ and $2\times 2$ enlarged unit cells at $n=\frac{1}{3}$, $\frac{1}{4}$ respectively.

In the LLL, the formation of charge density wave states is inevitable when the low filling factor is sufficiently low or, equivalently, the ratio between the average inter-particle spacing and the magnetic length is sufficiently large. While the precise filling factor below which such states are favored is not firmly agreed upon, it is widely believed to be around $\nu\approx \frac{1}{7}$ \cite{lam1984liquid,yang2001wigner}. Moreover, particle-hole symmetry in the LLL ensures that if a CDW of particles is the ground state at filling factor $\nu$, a CDW of holes is the ground state at $1-\nu$. In topological bands of $t$TMDs, the presence of CDWs at relatively high filling factors, and the lack thereof at their particle-hole conjugates, illustrates a clear departure of the system at $\theta>\theta_m$ from the LLL.

The nature of the $n=\frac{1}{3}$ FQAH-CDW and $n=\frac{1}{4}$ ACFL-CDW phase transitions are interesting questions for future theoretical investigation. Note that the CDW calculations presented here assume full spin/valley polarization, which can be achieved by applying a small magnetic field. We leave a systematic study of the magnetism of CDW states at zero field to future work. 


\section{Discussion and Outlook}

In this work, we have begun to map out a global phase diagram of the fractional quantum anomalous Hall effect that occurs in partially filled topological bands of $t$TMDs. With an exact diagonalization study, we have shown that near a magic angle $\theta_m$ the phase diagram bears remarkable resemblance to that of the LLL, hosting anomalous composite Fermi liquid phases at $n=\frac{1}{2}$, $\frac{1}{4}$, $\frac{3}{4}$ and FQAH phases at Jain sequence filling fractions $n=\frac{1}{3}$, $\frac{2}{5}$, $\frac{3}{7}$, $\frac{4}{7}$, $\frac{3}{5}$, and $\frac{2}{3}$. 

Above the magic angle, we find the phase diagram to be strongly particle-hole asymmetric. 
Charge density waves appear at $n=\frac{1}{3}$, $\frac{1}{4}$. In contrast, anomalous Hall Fermi liquids appear at $\frac{2}{3} < n < 1$. 
We find that the ACFL state at half-filling is particularly robust against deviations from the magic angle, surviving to even larger angles than $n=\frac{2}{3}$ FQAH state. Our phase diagram explains the recent observation of (1) an incompressible trivial state at $n=\frac{1}{3}$ and (2) filling-dependent anomalous Hall effect in the metallic region $\frac{2}{3} < n < 1$ in $t$MoTe$_2$ devices that show an $n=\frac{2}{3}$ FQAH state. 

Recent many-body calculations using different continuum model parameters \cite{wang2023fractional,dong2023composite} find FQAH states at both $n=\frac{1}{3}$ and $\frac{2}{3}$ in the twist angle range $\theta_m \sim 3-4^{\circ}$ where the majority of experiments to date have been performed. In contrast, our work finds that the simultaneous appearance of these FQAH states is limited to the neighborhood of a magic angle, $\theta_m \sim 2^{\circ}$. While there is uncertainty about the continuum model parameters and, therefore, the precise value of $\theta_m$, the appearance of a trivial insulating state at $n=\frac{1}{3}$ \cite{zeng2023thermodynamic}, consistent with a CDW, indicates that devices studied to date are above the magic angle. 

While the ED calculations presented in this work are performed specifically for $t$MoTe$_2$, we expect our main conclusions as stated above 
to hold qualitatively for a broader class of Chern band systems with band-projected Hamiltonians that can be approximately mapped to that of the LLL. These include Chern bands from periodically modulated Landau levels \cite{dong2022dirac}, skrymion lattices in semiconductor-magnet heterostructures \cite{paul2023giant}, and graphene moir\'e superlattices \cite{zhang2019nearly,repellin2020chern,abouelkomsan2020particle,ledwith2020fractional,wilhelm2021interplay,parker2021field} which host fractional Chern insulator states at strong magnetic field \cite{spanton2018observation,xie2021fractional}.

The discovery of integer and fractional quantum Hall effects in the two-dimensional electron gas at high magnetic field ushered in a revolution of topological quantum physics that remains fruitful over forty years later \cite{tsui1982two}. The possibility of realizing analogous topological quantum fluids in Chern bands, broadly referred to as Chern insulators and fractional Chern insulators, was envisioned and demonstrated by proof-of-principle lattice model studies \cite{haldane1988model,sun2011nearly,tang2011high,neupert2011fractional,sheng2011fractional,regnault2011fractional}.

Of particular interest and fundamental importance is the quantized anomalous Hall effect at zero magnetic field \cite{chang2023colloquium}, which may occur in time-reversal-invariant topological bands \cite{wu2019topological,song2015topological,kane2005quantum} and requires magnetic ordering that {\it spontaneously} breaks time reversal symmetry \cite{sharpe2019emergent, serlin2020intrinsic, devakul2021magic,crepel2023anomalous,reddy2023fractional,morales2023pressure}. Thanks to innovation in moiré quantum materials and advanced theoretical guidance, the fractionally quantized anomalous Hall effect has finally been observed \cite{cai2023signatures,zeng2023thermodynamic, park2023observation}. In the same spirit as a partially-filled Landau level, a partially-filled topological band can exhibit a symphony of distinct phases as a function of filling factor, each bringing its own novelty 
as an impetus to extend the frontier of condensed matter physics. 

Recent experiments on $t$MoTe$_2$ have demonstrated the ability to induce phase transitions out of QAH states by  applying displacement field \cite{cai2023signatures,zeng2023thermodynamic,park2023observation}. These phase transitions have received limited theoretical and numerical attention thus far and provide an interesting direction for future work. Pressure, an additional \emph{in situ} tuning knob that has been demonstrated numerically to enhance the stability of FQAH states, also deserves attention \cite{morales2023pressure}.

While our global phase diagram contains many prominent features supported both by our numerical study and/or experiment, it is by no means complete. Further study of the various phases we identify by other numerical methods with access to larger system sizes such as density matrix renormalization group can provide valuable insight and have indeed already done so in the case of the ACFL \cite{dong2023composite}. Theoretical studies indicate that the IQAH state at $n=1$ may be energetically outcompeted at larger angles \cite{qiu2023interaction,wang2023topological}, and the possibility of competing phases at fractional filling not captured in the single band-projected model also deserves further investigation \cite{xu2023maximally}.

\emph{Acknowledgements--} We thank Hart Goldman, Nisarga Paul, Ahmed Abouelkomsan, and Emil Bergholtz for related collaborations. This work is supported by the Air Force Office
of Scientific Research (AFOSR) under Award No. FA9550-22-1-0432 and the Simons Investigator award from the Simons Foundation. The authors acknowledge the MIT SuperCloud and Lincoln Laboratory Supercomputing Center for providing HPC resources that have contributed to the research results reported within this paper.

\clearpage

\end{document}


\title{Supplemental material for \emph{{Toward a  global phase diagram of the fractional quantum anomalous Hall effect}}}
\author{Aidan P. Reddy}
\author{Liang Fu}
\affiliation{Department of Physics, Massachusetts Institute of Technology, Cambridge, Massachusetts 02139, USA}
\date{\today}
\date{\today}
\maketitle
\tableofcontents

\section{Extended exact diagonalization data}
Here we present numerical data to further characterize the quantum many-body phase diagram of $t$MoTe$_2$. These results illustrate $t$MoTe$_2$'s resemblance to the LLL near the magic angle and its departure from the LLL with increasing twist angle. All calculations here assume full spin/valley polarization. While experiments \cite{anderson2023programming,cai2023signatures,zeng2023thermodynamic} and previous numerical studies \cite{crepel2023anomalous,reddy2023fractional,morales2023magic,goldman2023zero} show evidence for full valley polarization throughout a broad range of filling factors and twist angles, it should not be taken for granted that the true ground state is fully spin-polarized under all conditions studied here. As in the main text, a coulomb interaction $V(\bm{r})=\frac{e^2}{\epsilon r}$ is used throughout.

\begin{figure*}
    \centering
\includegraphics[width=0.8\textwidth]{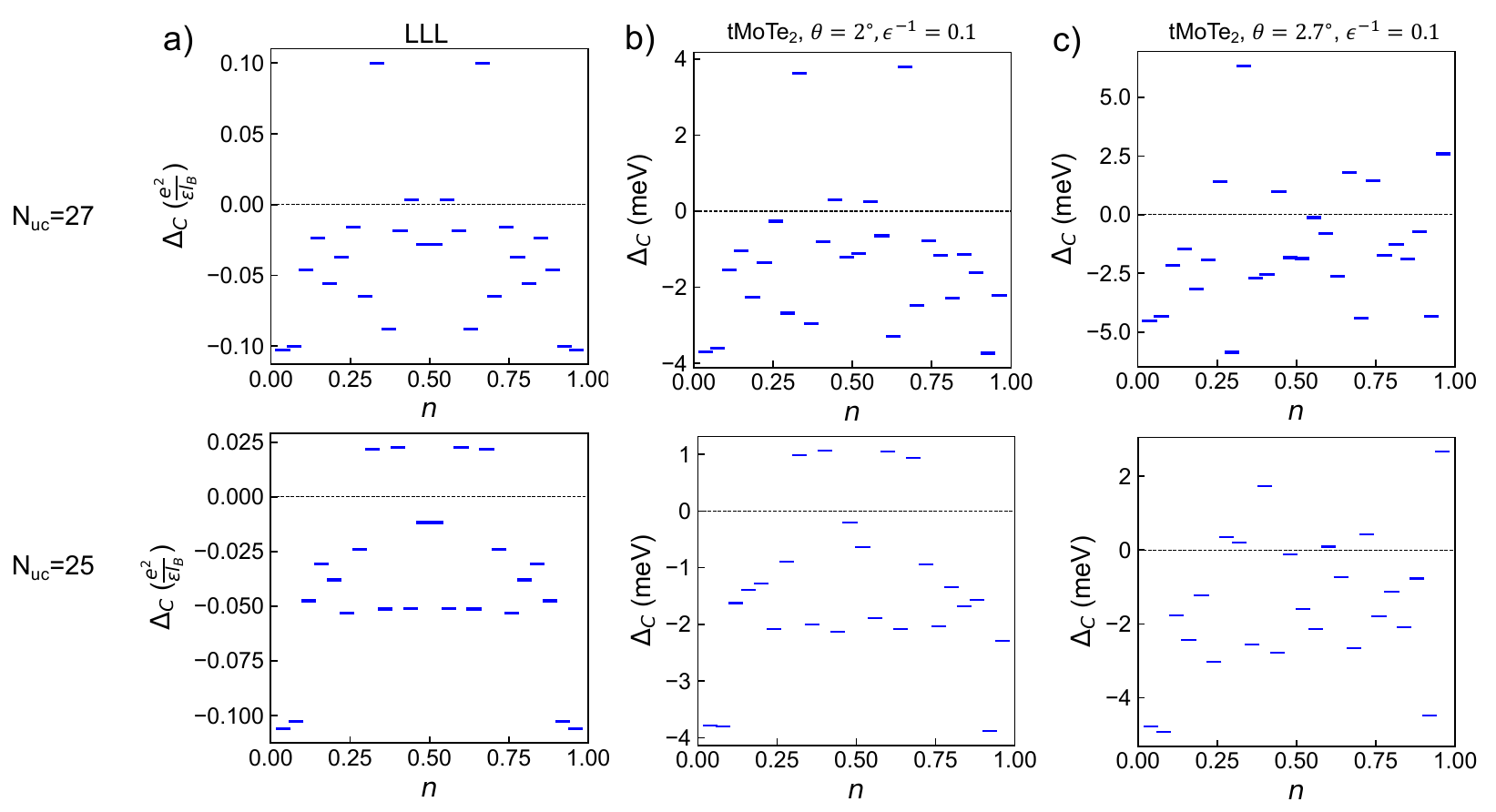}
    \caption{Finite size charge gap of the LLL (a) compared with that of $t$MoTe$_2$ near (b) and above (c) the magic angle on systems with 27 (top row) and 25 (bottom row) moiré unit cells. } 
\label{fig:chargegap}
\end{figure*}

\subsection{Charge gap}
First, in Fig. \ref{fig:chargegap} we show the finite-size charge gap ($\Delta_c(N)=\mu^{+}_{N}-\mu^{-}_{N}$ with $\mu^{\pm}_{N}=\pm(E_{GS}(N\pm1)-E_{GS}(N))$ where $N$ is the number of holes and $E_{GS}(N)$ is the ground state energy at a given $N$ and a fixed finite system geometry) of $t$MoTe$_2$ near and above the magic angle as a function of filling factor $n$ (= number of holes per moiré unit cell) alongside the corresponding data of the LLL. We caution that, by definition, compressible states have vanishing charge gaps in the thermodynamic limit. Therefore, the finite-size charge gap of such states is \emph{not} reflective of the thermodynamic limit. Our intent in presenting this data is simply to make the following qualitative points. Near the magic angle, the $t$MoTe$_2$ charge gap is nearly particle-hole symmetric and its evolution with $n$ closely resembles that of the LLL. Above the magic angle, the charge gap is strongly particle-hole asymmetric.

\begin{figure*}
    \centering
\includegraphics[width=\textwidth]{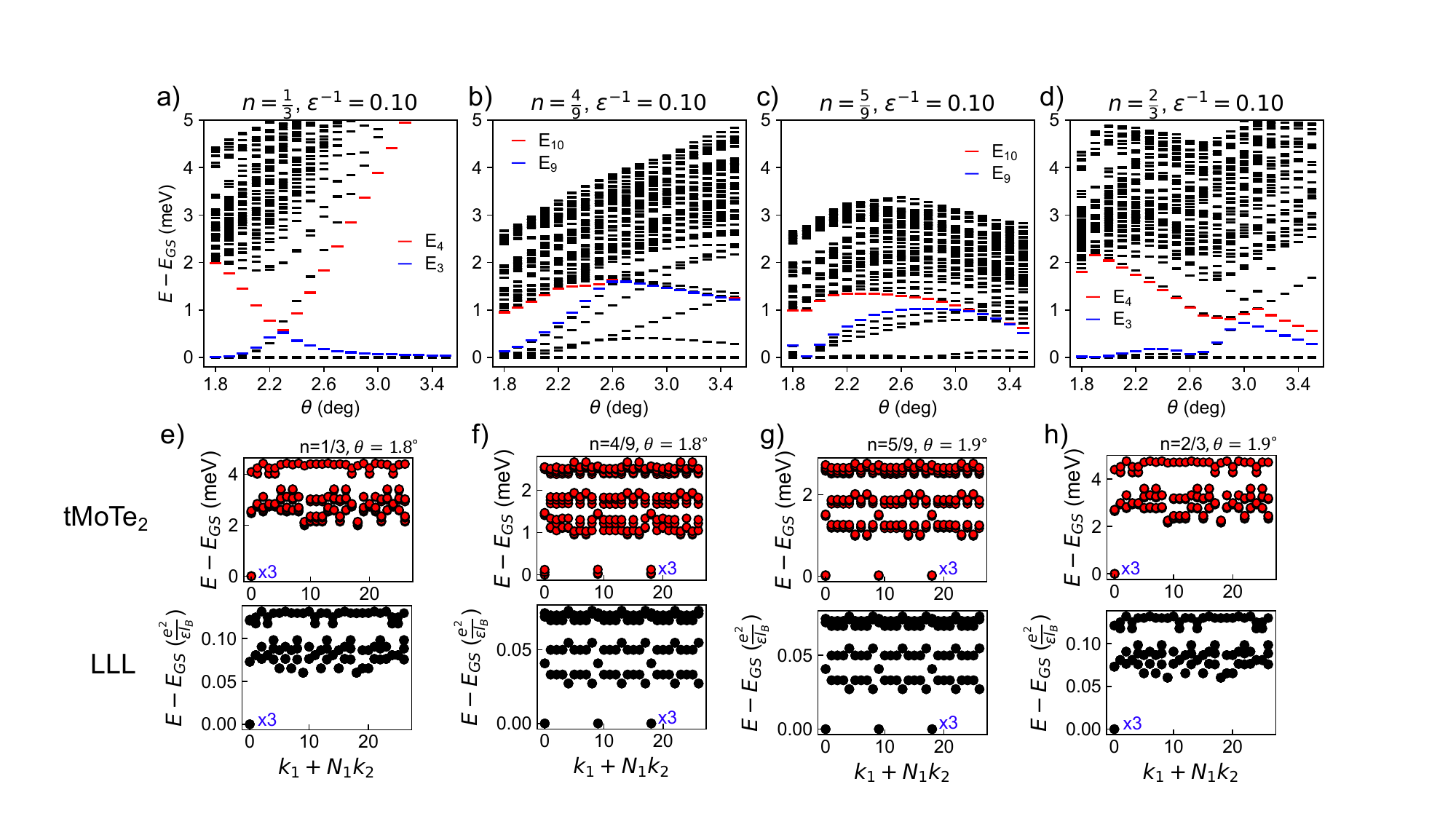}
    \caption{Low-lying many-body spectra as a function of twist angle at several filling fractions (a-d). Comparison of low-lying many-body spectra of $t$MoTe$_2$ near the magic angle with corresponding data of the LLL. The 10 lowest energy levels in each many-body momentum sector are shown. A cluster with 27 unit cells is used and $\epsilon^{-1}=10$ are used throughout. (See $s(\bm{q})$ inset to Fig. 2(c,d) of the main text for an illustration of the corresponding Brillouin zone mesh.)} 
\label{fig:27spectrum}
\end{figure*}

\begin{figure*}
    \centering
\includegraphics[width=\textwidth]{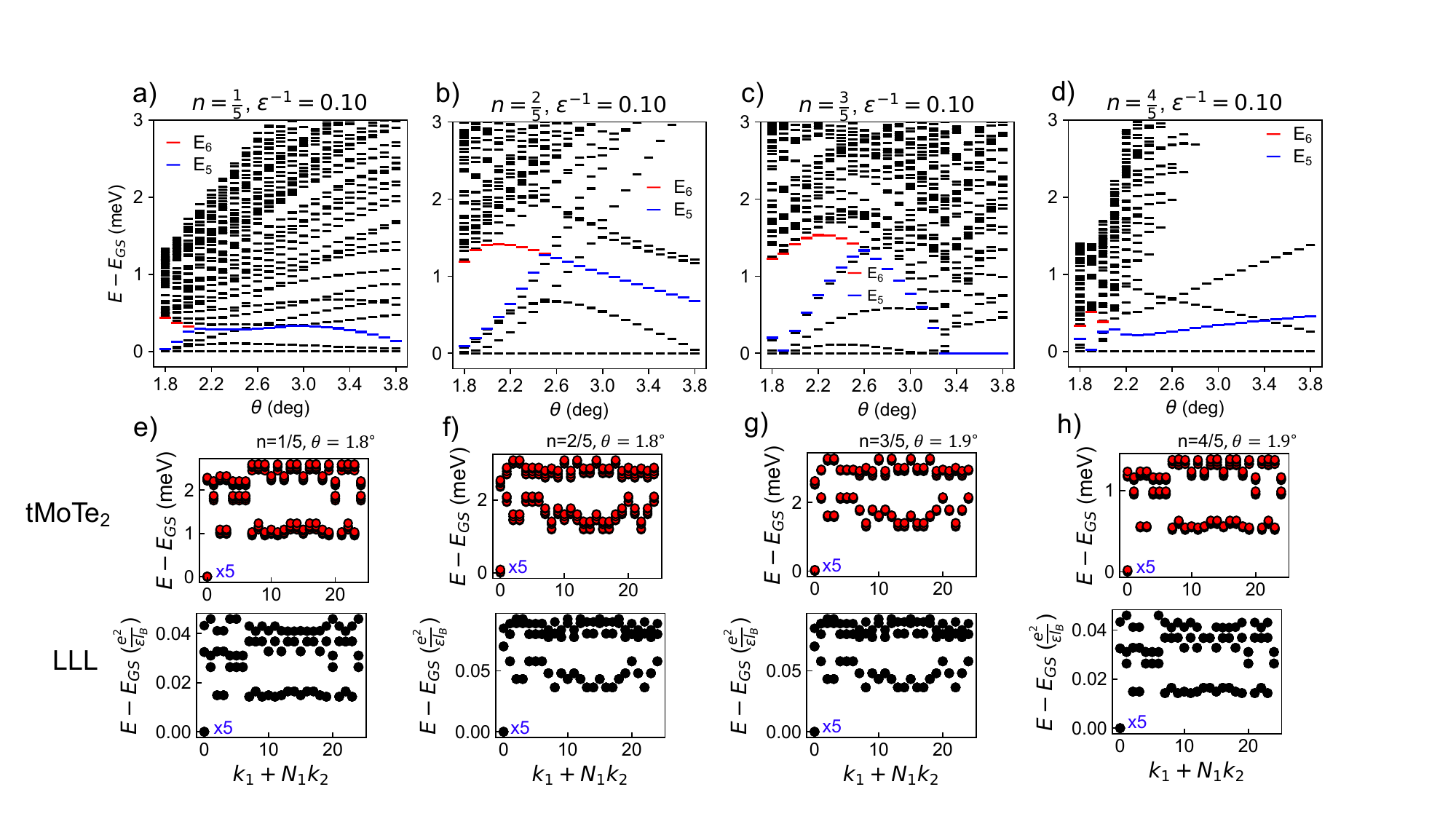}
    \caption{Similar to Fig. \ref{fig:27spectrum}, except on a cluster with 25 unit cells.}
\label{fig:25spectrum}
\end{figure*}

\subsection{Many-body spectra and comparison to LLL}
\begin{figure*}
    \centering
\includegraphics[width=0.9\textwidth]{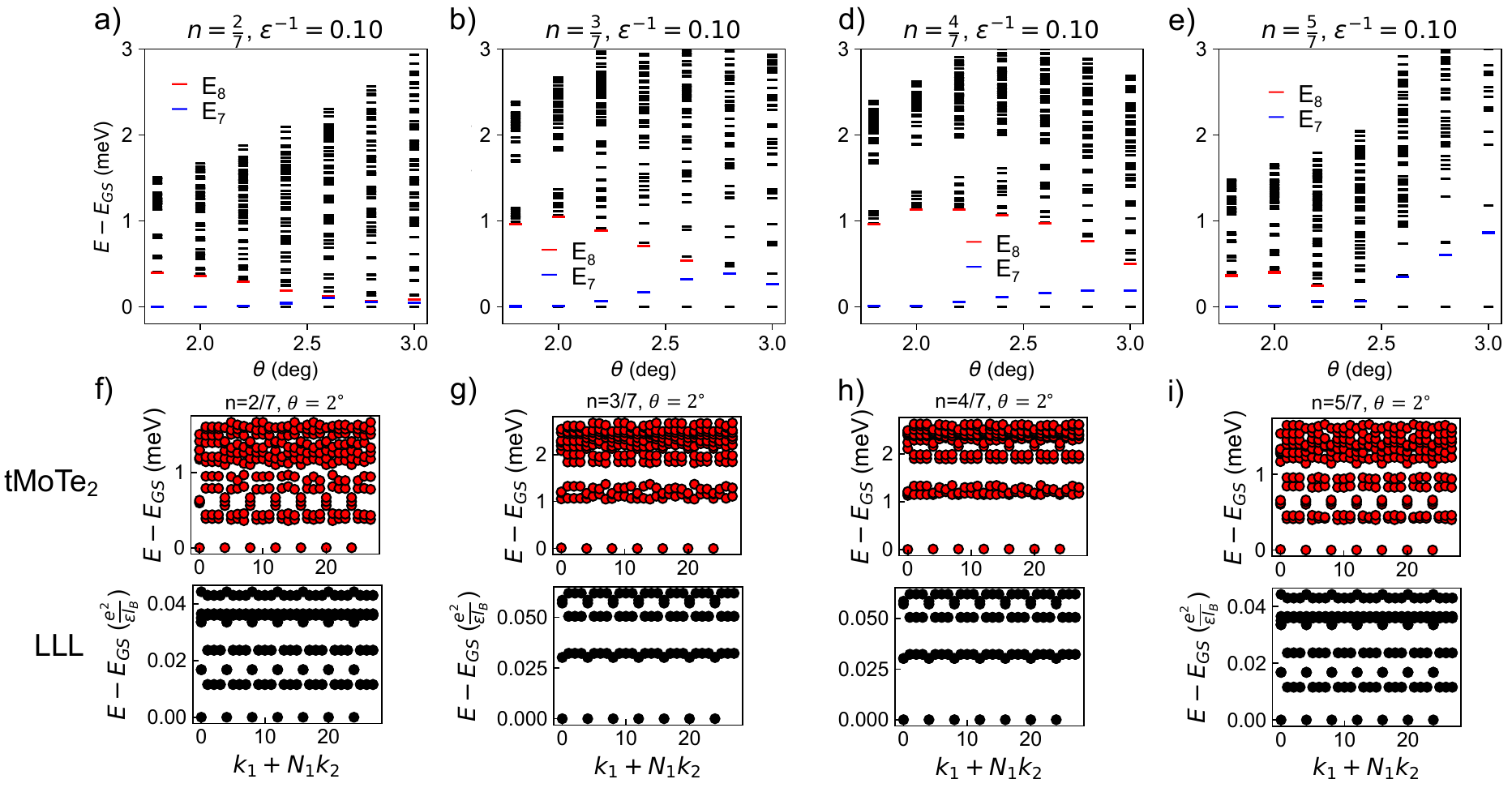}
    \caption{Similar to Fig. \ref{fig:27spectrum}, except on a cluster with 28 unit cells (see Fig. 3 of the main text for an illustration of the Brillouin zone mesh).} 
\label{fig:28spectrum}
\end{figure*}

Next, in Figs. \ref{fig:27spectrum},\ref{fig:25spectrum}, and \ref{fig:28spectrum} we plot the low-lying many-body spectra at several filling factors as a function of twist angle. In each case, near the magic angle, a $q$-fold quasi-degenerate ground state manifold is separated from excited states by a neutral (= fixed particle number) energy gap where $n=\frac{p}{q}$ is the filling fraction. By direct comparison of these states' many-body spectra to those of the LLL at identical filling fractions and system geometries, we confirm their FQAH nature. As twist angle increases, the neutral FQAH gap, $E_{gap}\equiv E_{q+1}-E_{q}$ where $E_{q}$ is $q^{th}$ lowest-energy state counting from 1, eventually closes. $E_{gap}$ passing through zero approximately indicates a phase transition out of the FQAH state. In general, the gap closing occurs at lower twist angles for filling fractions near $n=0$, $1$ and at higher twist angles for filling fractions near $n=\frac{1}{2}$. The enhanced stability against deviations from the magic angle of FQAH states near $n=\frac{1}{2}$ is suggestive of the $n=\frac{1}{2}$ anomalous composite Fermi liquid's role as a parent state.

\begin{figure*}
    \centering
\includegraphics[width=\textwidth]{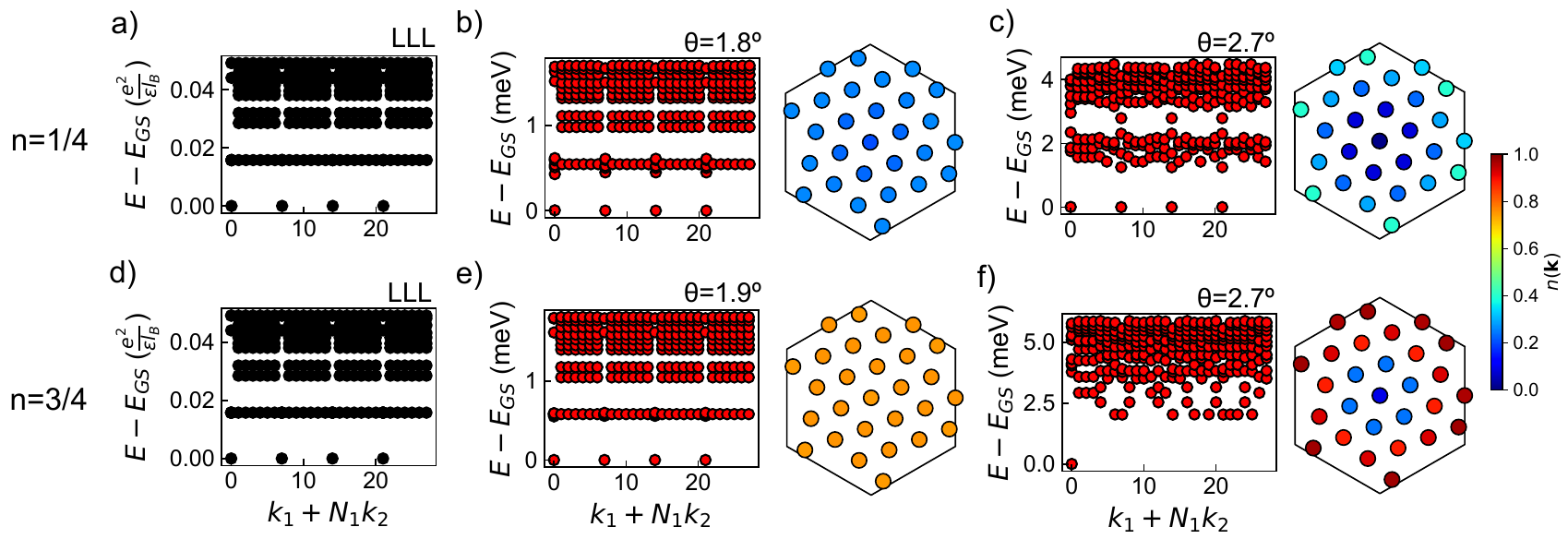}
    \caption{Many-body spectra of the LLL at $n=\frac{1}{4}$ (a) and $\frac{3}{4}$ (d) compared with that of $t$MoTe$_2$ near (b,e) and above (c,f) the magic angle. (b,c,e,f) also corresponding momentum distribution function. The lowest ten eigenvalues in each momentum sector are shown. $\epsilon^{-1}=0.1$.} 
\label{fig:quarterCFL}
\end{figure*}

In Fig. \ref{fig:quarterCFL}, we show many-body spectra and momentum distribution functions at filling factors $n=\frac{1}{4}$, $\frac{3}{4}$. Near the magic angle, the many-body spectra resemble that of the LLL and the momentum distribution functions do not exhibit Fermi surfaces. At larger angle, $n=\frac{1}{4}$ has four nearly degenerate ground states with many-body momenta at $\gamma$ and the three $m$ points of the moiré Brillouin zone. This is consistent with our diagnosis of this state in the main text based on its pair correlation function as a charge density wave with a $2 \times 2$ enlarged unit cell. In contrast, $n=\frac{3}{4}$ above the magic angle exhibits the characteristics of an anomalous Hall Fermi liquid state: a sharp Fermi surface and a ground state degeneracy consistent with finite-size ``shell filling". In this case, $7$ electron quasiparticles fill the Bloch states closest to the Brillouin zone center in a closed-shell configuration, leading to a unique ground state.

\begin{figure*}
    \centering
\includegraphics[width=0.6\textwidth]{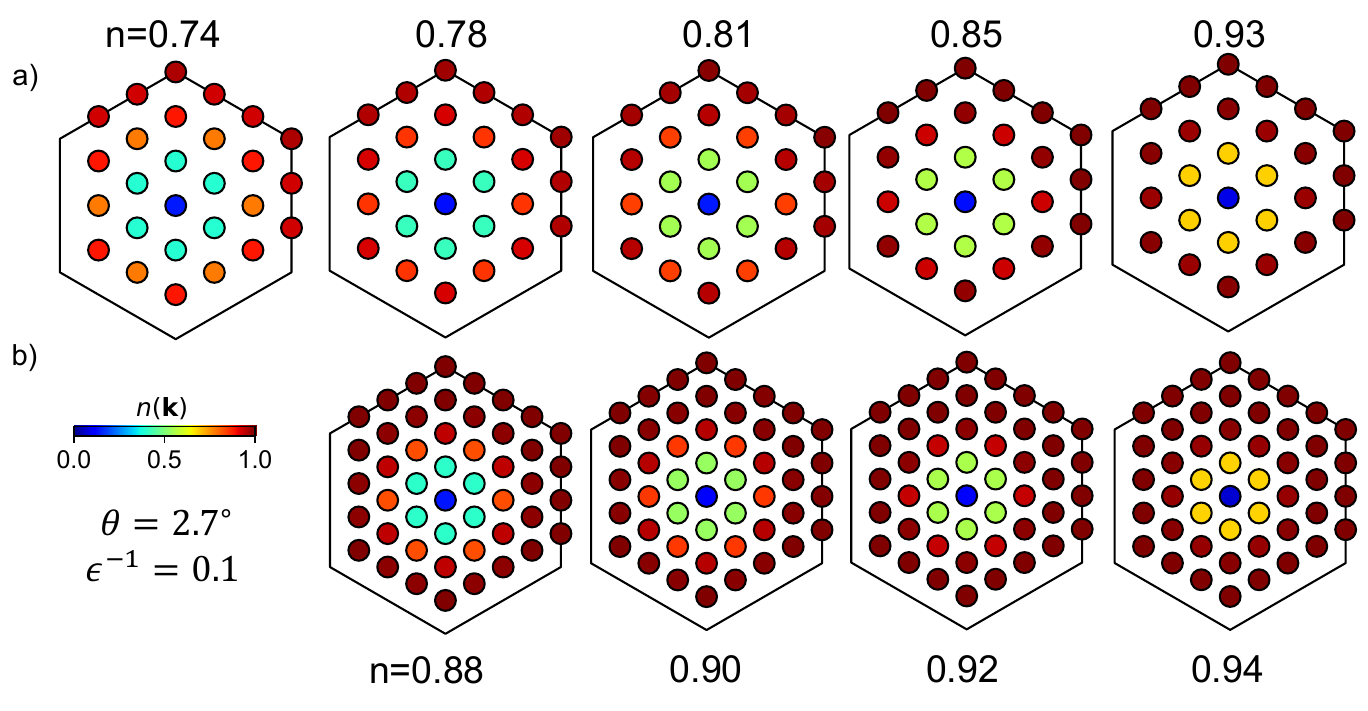}
    \caption{Momentum distribution function $n(\bm{k})$ at several filling factors $\frac{2}{3}<n<1$ above the magic angle on clusters with 27 (a) and 48 (b) unit cells.} 
\label{fig:AHFL}
\end{figure*}


In Fig. \ref{fig:AHFL}, we show momentum distribution function $n(\bm{k})$ data from additional system sizes and filling factors in the anomalous Hall Fermi liquid phase, which occurs above the magic angle and in the filling range $\frac{2}{3}<n<1$. $n(\bm{k}) = \langle c_{\bm{k}}^{\dag}c_{\bm{k}} \rangle$ where $c^{\dag}_{\bm{k}}$ creates a hole and $\langle \,\rangle$ denotes an average over degenerate ground states. In each case, a finite-size ``Fermi surface", across which the momentum distribution function drops sharply, is apparent.

\begin{figure*}
    \centering
\includegraphics[width=0.25\textwidth]{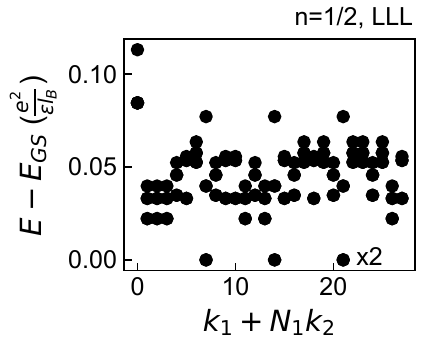}
    \caption{Many-body spectrum at $n=\frac{1}{2}$ of the LLL on a system with 28 flux quanta. (For comparison with Fig. 3 of the main text).} 
\label{fig:CFLLLL}
\end{figure*}

Finally, for comparison with Fig. 3 of the main text, we show the spectrum of the $n=\frac{1}{2}$ composite Fermi liquid in the LLL in Fig. \ref{fig:CFLLLL}.

\subsection{Correlation functions}

\begin{figure*}
    \centering
\includegraphics[width=0.65\textwidth]{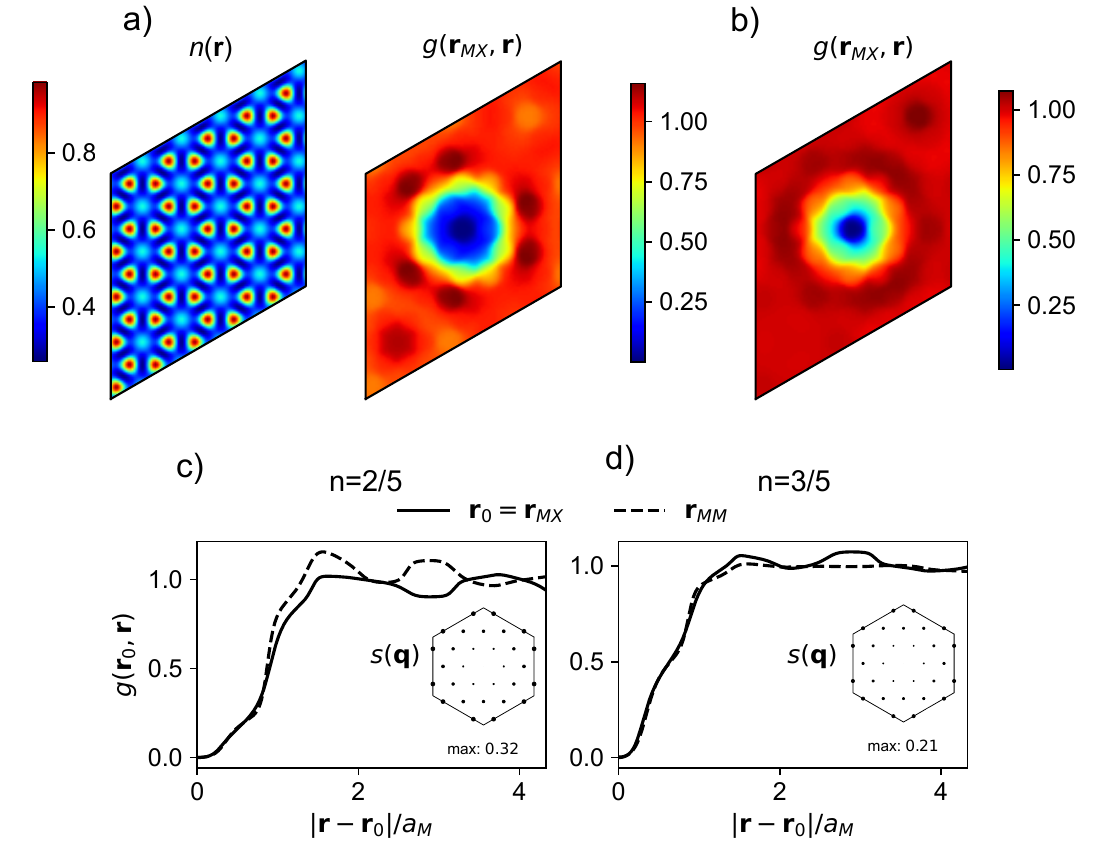}
    \caption{(a) Charge density and (b) pair correlation function $g(\bm{r},\bm{r}')$ of $t$MoTe$_2$ near the magic angle at $n=\frac{2}{5}$.  (b) Pair correlation function at $n=\frac{3}{5}$. (c,d) Show corresponding line cuts. Projected static structures $s(\bm{q})$ are shown as insets. $\theta=2^{\circ}$, $\epsilon=5$. System size is 25 moiré unit cells.} 
\label{fig:5paircorr}
\end{figure*}

In Fig. \ref{fig:5paircorr}, we show the pair correlation functions and projected static structure factors (see below for definition) of the $n=\frac{2}{5}$, $\frac{3}{5}$ FQAH states. Both states exhibit exchange-correlation holes at short distances and liquid-like correlations at intermediate and longer distances. Similarly to that of the $n=\frac{1}{3}$ state shown in the main text, the pair correlation function of the $\frac{2}{5}$ state exhibits oscillations indicative of incipient crystalline order. 

\subsection{Two-body spectrum}

The two-body interaction spectrum of band-projected Hamiltonians has been studied as an indicator of their likelihood to host FQAH states \cite{liu2013non,lauchli2013hierarchy,wang2021exact}. The different many-body phases appearing in different LL's can be understood in terms of differences in their two-body interaction spectra (=Haldane pseudopotentials \cite{haldane1983fractional}). Due to quenching of kinetic energy in a LL, the energy is independent of the center-of-mass motion, leading to large degeneracy in the two-body spectrum \cite{girvin2019modern}. In contrast to the higher LL's, the LLL exhibits a large gap between its highest and second-highest energy two-body states. From the perspective of hard-core pseudopotential Hamiltonians for Laughlin states \cite{trugman1985exact,macdonald1994introduction}, this property is responsible for the stability of its FQH states. 

\begin{figure*}
    \centering
\includegraphics[width=0.6\textwidth]{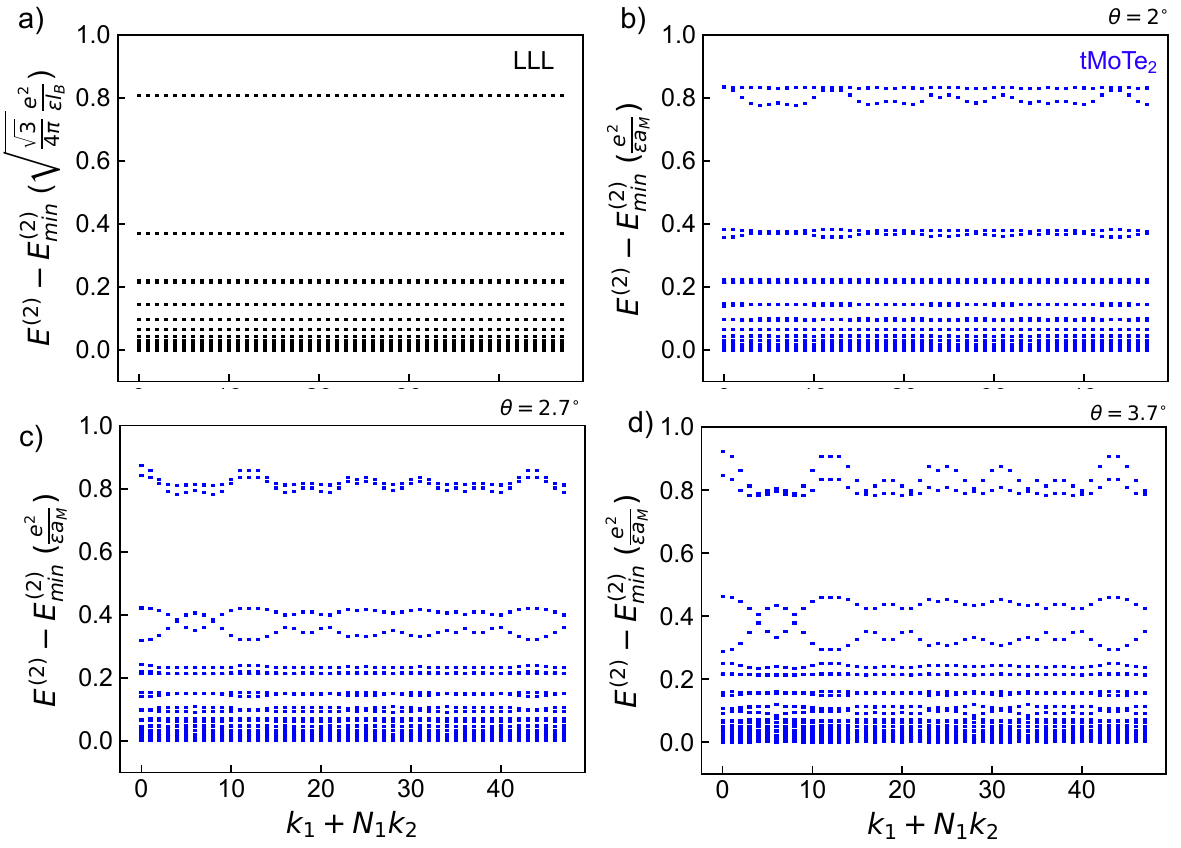}
    \caption{Two-body spectrum (neglecting bandwidth) of $t$MoTe$_2$ at several twist angles compared to that of the LLL at fixed 48 moiré unit cell system geometry. See Fig. 6 for $\bm{k}$-space mesh.} 
\label{fig:twobody}
\end{figure*}

In Fig. \ref{fig:twobody}, we show two-body interaction spectra for the lowest $t$MoTe$_2$ moiré band (with band dispersion turned off) at several twist angles alongside that of the LLL. The resemblance near the magic angle $\theta_m\approx 2^{\circ}$ is clear. The massive quasidegeneracy of the two-body spectrum is consistent with an ``emergent magnetic translation symmetry" \cite{bernevig2012emergent} and with all the other signatures of LLL-like behavior demonstrated in this work. At larger angles, the spectrum depends more strongly on the center-of-mass crystal momentum quantum number.

\section{Methods}

A discussion of the continuum model for $t$MoTe$_2$ and of our band-projected exact diagonalization method is provided in Ref. \cite{reddy2023fractional} and its Supplemental Material. For the LLL calculations, we use the methods described in the Supplemental Material of Ref. \cite{wang2021exact}.

\subsection{Calculating correlation functions}

Here we discuss how to calculate the two-body correlation functions studied in this work, the pair correlation function and projected structure factor. Throughout this section, we neglect spin. We define the pair correlation function as
\begin{align}
    g(\bm{r},\bm{r}') = \frac{\langle n(\bm{r})n(\bm{r}')\rangle - \delta(\bm{r}-\bm{r}')\langle n(\bm{r})\rangle}{\langle n(\bm{r})\rangle\langle n(\bm{r}')\rangle}.
\end{align}
The hole density operator is
\begin{align}
    n(\bm{r}) = \sum_{l}\psi_l^{\dag}(\bm{r})\psi_l(\bm{r})
\end{align}
where $l$ is a layer index. Therefore,
\begin{align}
    \begin{split}
        g(\bm{r},\bm{r}') &= \frac{\langle \sum_{l,l'}\psi_l^{\dag}(\bm{r})\psi_l(\bm{r})\psi_{l'}^{\dag}(\bm{r}')\psi_{l'}(\bm{r}')\rangle - \delta(\bm{r}-\bm{r}') \langle \sum_{l}\psi_l^{\dag}(\bm{r})\psi_l(\bm{r})\rangle }{\langle n(\bm{r})\rangle\langle n(\bm{r}')\rangle} \\
        &= \frac{\langle \sum_{l,l'}\psi_l^{\dag}(\bm{r})[\{\psi_l(\bm{r}),\psi_{l'}^{\dag}(\bm{r}')\} - \psi_{l'}^{\dag}(\bm{r}')\psi_l(\bm{r})]\psi_{l'}(\bm{r}')\rangle - \delta(\bm{r}-\bm{r}') \langle \sum_{l}\psi_l^{\dag}(\bm{r})\psi_l(\bm{r})\rangle }{\langle n(\bm{r})\rangle\langle n(\bm{r}')\rangle} \\
        &= \frac{\langle \sum_{l,l'}\psi_l^{\dag}(\bm{r})[\delta_{ll'} \delta(\bm{r}-\bm{r}') - \psi_{l'}^{\dag}(\bm{r}')\psi_l(\bm{r})]\psi_{l'}(\bm{r}')\rangle - \delta(\bm{r}-\bm{r}') \langle \sum_{l}\psi_l^{\dag}(\bm{r})\psi_l(\bm{r})\rangle }{\langle n(\bm{r})\rangle\langle n(\bm{r}')\rangle} \\
        &= \frac{\langle \sum_{l,l'}\psi_{l'}^{\dag}(\bm{r}')\psi_l^{\dag}(\bm{r})\psi_l(\bm{r})\psi_{l'}(\bm{r}')\rangle}{\langle n(\bm{r})\rangle\langle n(\bm{r}')\rangle}. 
    \end{split}
\end{align}
The field operators can be written in terms of Bloch state hole creation and annihilation operators as
\begin{align}
    \begin{split}
        \psi^{\dag}_l(\bm{r}) &= \sum_{n,\bm{k}} \bra{n,\bm{k}}\ket{\bm{r},l}c^{\dag}_{n\bm{k}} \\
        &\equiv \sum_{n,\bm{k}} \psi^*_{n\bm{k}l}(\bm{r})c^{\dag}_{n\bm{k}}
    \end{split}
\end{align}
where $\bm{k}$ is a moiré crystal momentum vector and $n$ is a moiré band index. Finally, exploiting crystal momentum conservation, we have
\begin{align}
    \begin{split}
        g(\bm{r},\bm{r}') &= \sum_{\bm{k},\bm{k}',\bm{p}} \langle c^{\dag}_{[\bm{k}+\bm{p}]}c^{\dag}_{[\bm{k}'-\bm{p}]}c_{\bm{k}'}c_{\bm{k}}\rangle \left(\frac{\sum_{l,l'} \psi^*_{[\bm{k}+\bm{p}]l}(\bm{r'})\psi^*_{[\bm{k}'-\bm{p}]l'}(\bm{r})\psi_{\bm{k}'l'}(\bm{r})\psi_{\bm{k}l}(\bm{r}')}{\langle n(\bm{r})\rangle\langle n(\bm{r}')\rangle} \right).
    \end{split}
\end{align}
Here, we assume (for simplicity and as pertains to this work) that the many-body state lives in the Fock space of a single band whose index we have dropped. The sum over $\bm{k},\bm{k}',\bm{p}$ runs over moiré crystal momenta. $[\bm{q}]$ denotes a chosen unique ``mesh" representative of the crystal momentum equivalence class $\bm{q} \mod \bm{g}$ of an arbitrary wavector $\bm{q}$. The moiré Bloch state wavefunctions are
\begin{align}
    \begin{split}
        \psi_{n\bm{k}l}(\bm{r}) &\equiv \bra{\bm{r},l}\ket{n,\bm{k}} \\
        &= \bra{\bm{r},l}\left(\sum_{\bm{g}}\ket{\bm{k}+\bm{g},l}\bra{\bm{k}+\bm{g},l}\ket{n,\bm{k}}\right) \\
        &= \sum_{\bm{g},l} \bra{\bm{k}+\bm{g},l}\ket{n,\bm{k}}e^{i(\bm{k}+\bm{g})\cdot\bm{r}}.
    \end{split}
\end{align}
The Supplemental Material of Ref. \cite{reddy2023fractional} describes how to calculate the moiré Bloch states $\ket{n,\bm{k}}$.

The ``projected structure factor" is 
\begin{align}
\begin{split}
        s(\bm{q}) &= \frac{\langle \bar{\rho}(-\bm{q})\bar{\rho}(\bm{q})\rangle}{N}-\delta_{\bm{q},\bm{0}}
\end{split}
\end{align}
where $\bar{\rho}(\bm{q}) = \mathcal{P}\rho(\bm{r})\mathcal{P}$ is the projected density fluctuation operator and $\rho(\bm{q}) = \sum_{i}e^{-i\bm{q}\cdot\bm{r}_i}$ is the density fluctuation operator. $N$ is the total number of holes. $\mathcal{P}$ is a projector onto the Fock space of the lowest moiré band and $i$ indexes holes. The name ``projected structure factor" is somewhat misleading because $s(\bm{q}) \neq \mathcal{P}S(\bm{q})\mathcal{P}$ where $S(\bm{q}) = \frac{\langle \rho(-\bm{q})\rho(-\bm{q})\rangle}{N}-\delta_{\bm{q},\bm{0}}$ is the standard structure factor. The reason for the two functions' inequivalence is that a projector $\mathcal{P}$ appears in between the two density fluctuation operators in $s(\bm{q})$ but not in $S(\bm{q})$. We note that $s(\bm{q})$ is a function of $\bm{q}$ absolutely, not only $\bm{q} \mod \bm{g}$. In this work, however, we only plot $s(\bm{q})$ for $\bm{q}$ in the first moiré Brillouin zone because this suffices to demonstrate the presence or absence of crystallinity. We add the $-\delta_{\bm{q},0}$ term to the definition of $s(\bm{q})$ to eliminate a $\bm{q}=0$ peak of 1 that would otherwise be present.

In second quantization, the projected density fluctuation operator is 
\begin{align}
    \begin{split}
      \bar{\rho}(\bm{q}) &= \sum_{\bm{k}} \bra{[\bm{k}-\bm{q}]}e^{-i\bm{q}\cdot\bm{r}}\ket{\bm{k}}c^{\dag}_{[\bm{k}-\bm{q}]}c_{\bm{k}} 
    \end{split}
\end{align}
Here we neglect band indices because  we are projecting to a single band. Bloch states in that band are labeled only by a wavector. Layer plane wave basis states, in contrast, also have a layer index $l$. The projected structure factor is then
\begin{align}
    s(\bm{q}) &=\sum_{\bm{k}\bm{k}'} \bra{[\bm{k}+\bm{q}]}e^{+i\bm{q}\cdot\bm{r}}\ket{\bm{k}} \bra{[\bm{k}'-\bm{q}]}e^{-i\bm{q}\cdot\bm{r}}\ket{\bm{k}'} \langle c^{\dag}_{[\bm{k}+\bm{q}]}c_{\bm{k}}c^{\dag}_{[\bm{k}'-\bm{q}]}c_{\bm{k}'}\rangle - \delta_{\bm{q},\bm{0}}.
\end{align}
Finally, the ``momentum boost" matrix elements are
\begin{align}
    \begin{split}
        \bra{[\bm{k}+\bm{q}]}e^{+i\bm{q}\cdot\bm{r}}\ket{\bm{k}} &= \sum_{l',l}\sum_{\bm{g}',\bm{g}} \bra{[\bm{k}+\bm{q}]}\ket{[\bm{k}+\bm{q}]+\bm{g}',l'}\bra{[\bm{k}+\bm{q}]+\bm{g}',l'}e^{+i\bm{q}\cdot\bm{r}}\ket{\bm{k}+\bm{g},l}\bra{\bm{k}+\bm{g},l}\ket{\bm{k}} \\
        &= \sum_{l',l}\sum_{\bm{g}',\bm{g}} \bra{[\bm{k}+\bm{q}]}\ket{[\bm{k}+\bm{q}]+\bm{g}',l'}\delta_{l',l}\delta_{[\bm{k}+\bm{q}]+\bm{g}',\bm{k}+\bm{q}+\bm{g}}\bra{\bm{k}+\bm{g},l}\ket{\bm{k}} \\
        &= \sum_{l}\sum_{\bm{g}} \bra{[\bm{k}+\bm{q}]}\ket{\bm{k}+\bm{q}+\bm{g},l}\bra{\bm{k}+\bm{g},l}\ket{\bm{k}} \\
        &= \sum_{l}\sum_{\bm{g}} \bra{[\bm{k}+\bm{q}]}\ket{[\bm{k}+\bm{q}]+\bm{g}_{\bm{k}+\bm{q}}+\bm{g},l}\bra{\bm{k}+\bm{g},l}\ket{\bm{k}}
    \end{split}
\end{align}
where $\bm{g}_{\bm{q}} = \bm{q}-[\bm{q}]$.


\section{Hartree-Fock self-energy}

The subject of Hartree-Fock self-energy as it pertains to particle-hole transformations in Bloch band systems has been discussed extensively in the literature \cite{lauchli2013hierarchy,abouelkomsan2020particle,abouelkomsan2023quantum}. Explicit equations for the Hartree and Fock self energy terms used in the main text are:
\begin{align}
    \begin{split}
        \Sigma^{H}(\bm{k})&=\sum_{{\bm p}} V_{{\bm k}{\bm p} {\bm k} {\bm p}};\\
        \Sigma^{F}(\bm{k})&=-\sum_{\bm{p}}V_{{\bm k} {\bm p}{\bm p} {\bm k}}
    \end{split}
\end{align}
where the sum is over moiré crystal momenta $\bm{p}$ and $V_{\bm{kpqr}}=\bra{\bm{k},\bm{p}}\hat{V}\ket{\bm{q},\bm{r}}$ are two-body interaction matrix elements in the moiré Bloch state basis. An explanation of how to calculate the two-body interaction matrix elements is provided in the Supplemental Material of Ref. \cite{reddy2023fractional}.